\documentclass{article} 
\usepackage{iclr2027_conference,times}

\usepackage{amsmath,amsfonts,bm}

\def\eqref#1{equation~\ref{#1}}

\def\1{\bm{1}}

\DeclareMathAlphabet{\mathsfit}{\encodingdefault}{\sfdefault}{m}{sl}
\SetMathAlphabet{\mathsfit}{bold}{\encodingdefault}{\sfdefault}{bx}{n}

\usepackage{wrapfig}
\usepackage[table]{xcolor}
\usepackage{graphicx}
\usepackage[ruled,linesnumbered]{algorithm2e}
\usepackage{hyperref}
\usepackage{booktabs}
\usepackage{multirow}
\usepackage{amssymb}
\usepackage{makecell}
\usepackage{url}
\usepackage{caption}
\usepackage{xcolor}
\usepackage[most]{tcolorbox}

\definecolor{iclrblue}{HTML}{3B5BDB}

\newtcolorbox{keyfinding}{
    colback=iclrblue!4!white,
    colframe=cyan!8,
    boxrule=0pt,
    leftrule=2pt,
    arc=0pt,
    outer arc=0pt,
    left=10pt,
    right=4pt,
    top=4pt,
    bottom=4pt,
    before skip=8pt,
    after skip=8pt,
    fontupper=\color{black!85},
    before upper={
        \textbf{\color{iclrblue!85!black}Key Finding:}\quad
    }
}
\definecolor{analysisHypothesisFrame}{HTML}{D9E4F0}
\definecolor{analysisHypothesisBackground}{HTML}{F8F8F8}
\definecolor{analysisHypothesisText}{HTML}{004C99}
\tcbset{analysis hypothesis/.style={
    colback=analysisHypothesisBackground,
    colframe=analysisHypothesisFrame,
    colbacktitle=analysisHypothesisFrame,
    coltitle=black,
    coltext=analysisHypothesisText,
    fonttitle=\sffamily\bfseries\small,
    fontupper=\normalsize,
    boxrule=0.5pt,
    arc=2mm,
    boxsep=0pt,
    left=8pt, right=8pt, top=6pt, bottom=6pt,
    toptitle=4pt, bottomtitle=4pt,
    before skip=6pt, after skip=6pt,
    width=\linewidth
}}
\title{From Attention to Semantics: Detecting and Mitigating Sound Event Hallucinations in Large Audio-Language Models}

\author{
Tianyi Sun$^{1}$, Minghao Lv$^{1}$, Chunyang Yu$^{2}$, Shenyue Wang$^{2}$, Mengyue Wu$^{1}$ \\
$^{1}$Shanghai Jiao Tong University \\
$^{2}$Guangdong OPPO Mobile Telecommunications Corp., Ltd. \\
}
\iclrfinalcopy 
\begin{document}

\maketitle
\lhead{Preprint.}

\begin{abstract}
Large Audio-Language Models (LALMs) excel at diverse audio tasks, yet remain prone to hallucinating sound events absent from the input. Prior work links such hallucinations to insufficient audio attention. However, we find that real and hallucinated events can receive comparable audio attention mass while attending to different audio positions. To examine whether the attended representations support the generated events, we adapt LatentLens to retrieve neighboring contextualized text representations. The retrieved contexts reveal semantic support for real events, but a semantic mismatch between hallucinated events and their attended audio representations. Directly masking the corresponding audio regions can suppress hallucinated events at the cost of real content. We are motivated by these findings to introduce an explainable and training-free \emph{detect-then-mitigate} framework for effectively hallucination detection and mitigation that aggregates semantic support and applies differential audio masking to protect regions associated with events predicted to be supported. Extensive experiments on three LALMs and two datasets demonstrate that our method significantly improves hallucination detection and hallucination mitigation performance.
\end{abstract}

\section{Introduction}

Large audio language models (LALMs) have extended language-based interaction to speech, environmental sounds, and music, enabling a wide range of complex cross-modal tasks within a unified framework~\citep{ghosh2026audio,xu2025qwen25omnitechnicalreport,wu2025step}. Despite these advances, LALMs remain prone to generating plausible descriptions of sound events that are absent from the input audio~\citep{kuan24_interspeech}. We refer to this issue as \emph{sound event hallucination}\citep{kuan24_interspeech,kuan2025can}, which significantly limits the applicability of LALMs in high-stakes scenarios that require highly reliable acoustic perception, such as autonomous driving and audio-based healthcare monitoring.

\begin{figure}[htbp]
    \centering
    \includegraphics[width=\linewidth]{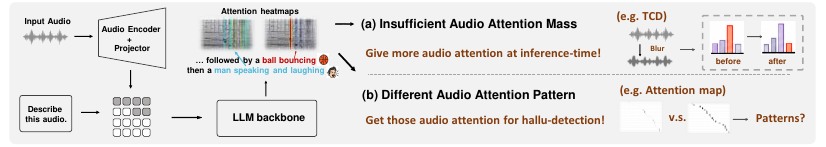}
    \caption{Previous efforts for audio hallucination detection or mitigation.}
    \label{fig:method-comparison}
\end{figure}

Prior empirical work links hallucination in LALMs to insufficient audio attention~\citep{seth2026audio}, suggesting that textual input can dominate acoustic evidence~\citep{wang2025audio}. Accordingly, many Inference-time methods such as AAD~\citep{hsu2025reducing} and TCD~\citep{li2026temporal}, as shown in Fig. ~\ref{fig:method-comparison}(a), seek to strengthen the contribution of audio evidence during decoding. Our findings, however, highlight a limitation of focusing on audio attention strength alone: real and hallucinated events can receive comparable audio attention mass. A similar observation has been reported in LVLMs, where real and hallucinated objects can receive equally strong visual attention~\citep{wang2026same,nguyen2026beyond}. Insufficient audio attention alone therefore does not explain sound event hallucination. This raises the question of where the model attends when the overall attention mass is comparable.

We next analyze the distribution of attention over audio tokens during the generation of real and hallucinated events. This analysis shows that attention can concentrate on different audio positions despite comparable total attention mass. This observation is consistent with prior findings that attention patterns provide useful signals for hallucination detection~\citep{waldendorf2026detecting}, as illustrated in Fig. ~\ref{fig:method-comparison}(b). We further ask whether these differently attended positions carry similar or distinct semantic information, and whether this information supports the generated event. 

To examine this semantic support, we interpret the audio representations most attended by each event. The representational gap between audio and text makes direct vocabulary projections difficult to interpret. Therefore, we adapt LatentLens~\citep{krojer2026latentlens} to retrieve contextualized text representations nearest to the attended audio representations. The retrieved contexts reveal a semantic distinction: they support real events, whereas hallucinated events exhibit a mismatch with the semantics retrieved from their attended representations.
Further, we test how intervening on regions associated with this semantic mismatch changes the output. Masking these regions can eliminate a hallucinated event, but can also remove real events from the regenerated response. This observation motivates protecting regions associated with supported events when selecting regions to mask.

Based on these findings, we propose a training-free \emph{detect-then-mitigate} framework for effective hallucination detection and targeted mitigation. For each generated sound event, we use LatentLens to estimate semantic support from its highly attended audio representations and flag events with insufficient support. We then apply \emph{differential audio masking}: positions attended by events predicted to be supported are protected, while unprotected positions associated with predicted hallucinations are mapped to audio regions and silenced before regeneration. Experiments on three LALMs and two datasets demonstrate improvements in hallucination detection and mitigation. Our contributions are summarized as follows:

\begin{itemize}
    \item We combine attention analysis with latent semantic retrieval to characterize sound event hallucinations in LALMs. Grounded and hallucinated events can receive comparable audio attention while attending to different positions, and hallucinated events exhibit a mismatch between the generated event and the semantics of the attended representations.

    \item We introduce a simple yet effective, explainable, and training-free framework for hallucination detection and mitigation that aggregates semantic support and applies differential audio masking to protect regions associated with events predicted to be supported.

    \item Extensive experiments on three LALMs and two datasets demonstrate that our method significantly improves hallucination detection and hallucination mitigation performance, validating its effectiveness and generalizability.
\end{itemize}

\vspace{-5pt}
\section{Related Work}

\vspace{-5pt}
\textbf{Large Audio-Language Models (LALMs).}
Large audio-language models (LALMs) connect audio encoders to language models via alignment modules, enabling open-ended understanding of speech, environmental sounds, and music. Early systems established complementary capabilities: Pengi cast audio tasks as text generation~\citep{deshmukh2023pengi}; LTU developed audio perception and reasoning via question-answer training~\citep{gong2024listen}; SALMONN integrated speech and general-audio encoders~\citep{tang2024salmonn}. Later models improved instruction following and interaction (e.g., Qwen2-Audio~\citep{chu2024qwen2}; Audio Flamingo~\citep{kong2024audioflamingo}). Recent work extends to long-audio captioning and reasoning in Audio Flamingo 2~\citep{pmlr-v267-ghosh25b}, unified speech/sound/music representations in Audio Flamingo 3~\citep{ghosh2026audio}, multimodal processing via Qwen2.5-Omni's Thinker--Talker architecture~\citep{xu2025qwen25omnitechnicalreport}, and spoken interaction in Step-Audio 2 mini~\citep{wu2025step}. However, aggregate performance does not show whether each generated sound event is supported by the input audio. We study event-level grounding across Audio Flamingo 3, Qwen2.5-Omni-7B, and Step-Audio 2 mini.

\textbf{Hallucination Detection and Mitigation in LALMs.}
Research on hallucinations in LALMs spans detection and mitigation. AudioLens traces auditory attributes via vocabulary projections of hidden states~\citep{yang2025audiolens}. Attention analyses link implicit hallucination attacks to reduced audio attention~\citep{seth2026audio}, and attention-derived features enable detection in speech recognition and translation~\citep{waldendorf2026detecting}. Mitigation includes training with synthesized negatives (LISTEN)~\citep{kuan25_interspeech} and counterfactual preference alignment (AHA)~\citep{chen2026aha}. Training-free methods contrast predictions with/without audio conditioning (AAD)~\citep{hsu2025reducing} or from original and temporally smoothed audio (TCD)~\citep{li2026temporal}. However, they do not resolve whether attended audio representations semantically support each generated event. We examine this event-level support and selectively mask audio positions of unsupported events while protecting those of supported events, linking semantic analysis of attended audio states to targeted intervention.

\section{Hallucination Analysis}
\label{sec:analysis}

\vspace{-5pt}
Why do large audio-language models generate sound events that are absent from the audio? Building on the prior empirical finding that insufficient attention to audio is one cause of hallucination~\citep{seth2026audio}, we first compare attention patterns between real and hallucinated events, then examine the semantic differences in the audio representations they attend to, and finally validate these mechanisms through targeted interventions. 

\vspace{-5pt}
\subsection{Preliminary}

\vspace{-5pt}
\textbf{LALM Generation Process.}  A general LALM comprises an audio encoder, a connector, and an LLM. Formally, an input audio clip is encoded into audio tokens \(A = \{a_1, \dots, a_N\}\) via the connector. These are then concatenated with system tokens \(S = \{s_1, \dots, s_M\}\) and text tokens \(T = \{t_1, \dots, t_K\}\) to form the LLM input \(I = \{s_1, \dots, s_M, a_1, \dots, a_N, t_1, \dots, t_K\}\). The LLM autoregressively generates a response \(C = \{c_1, \dots, c_L\}\), predicting each token as \(p(c_i \mid s_{1:M}, a_{1:N}, t_{1:K}, c_{<i})\).

\textbf{LatentLens.} LatentLens~\citep{krojer2026latentlens} is an interpretability method that interprets latent representations via nearest-neighbor retrieval over contextualized text representations. Given an audio-token hidden state \(h_t^{(\ell)}\), we compare it against a reference bank \(\mathcal{B}=\{(r_n,d_n)\}_{n=1}^{N}\), where \(r_n\) is a contextualized text-token representation from the same language model and \(d_n\) its associated textual context. Nearest neighbors are retrieved by cosine similarity:

\vspace{-5pt}
\[
\mathcal{N}_j\!\left(h_t^{(\ell)}\right)
=
\underset{n\in\{1,\ldots,N\}}{\operatorname{Top}\text{-}j}
\frac{\left(h_t^{(\ell)}\right)^\top r_n}
{\left\|h_t^{(\ell)}\right\|_2\left\|r_n\right\|_2}.
\tag{3}
\]
\vspace{-5pt}

The retrieved contexts \(\{d_n:n\in\mathcal{N}_j(h_t^{(\ell)})\}\) provide a textual interpretation of the queried representation. We apply LatentLens to highly attended audio-token hidden states to examine whether their semantics support the model's generated sound events.
\begin{figure*}[htbp]
    \centering
    \includegraphics[width=0.9\textwidth]{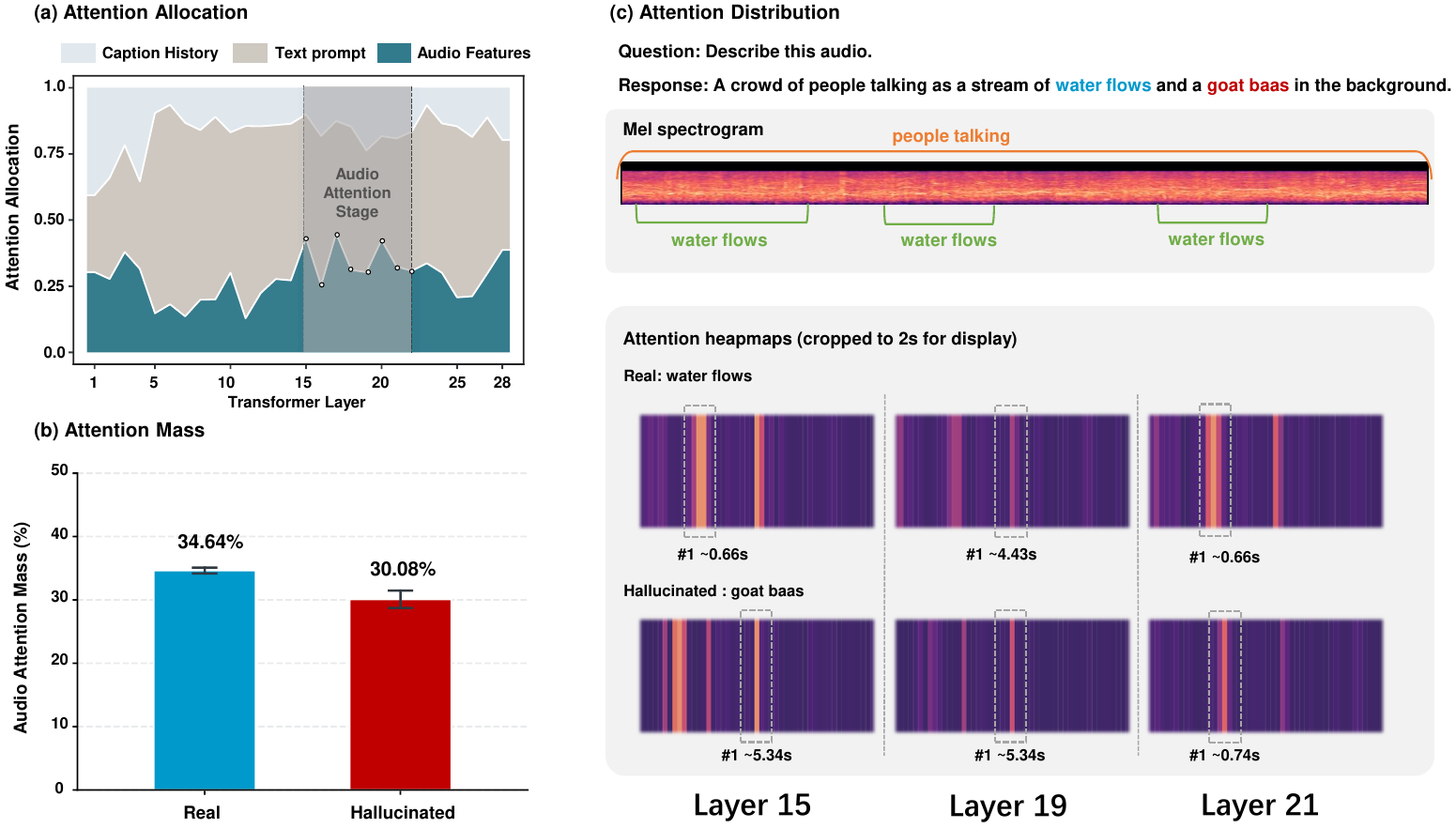}
    \caption{A typical case with audio attention analysis. (a) Attention allocation to audio features, text prompt, and caption history across layers. The shaded band marks the Audio Attention Stage (layers 15–22). (b) Mean audio attention mass for real and hallucinated events, averaged over the Audio Attention Stage; error bars denote 95\% bootstrap confidence intervals. (c) A representative response, its input Mel spectrogram with ground-truth sound event annotations, and attention maps for the real event ``water flows" and the hallucinated event ``goat baas" at layers 15, 19, and 21. Attention maps are cropped to 2-second windows; dashed boxes mark the most-attended audio positions, with approximate timestamps.} 
    \label{fig:audio-attention-analysis}
\end{figure*}

\vspace{-5pt}
\subsection{Comparable Audio Attention but Different Attended Positions}

\vspace{-5pt}
To illustrate, we analyze audio attention in Audio Flamingo 3~\citep{ghosh2026audio} on a set of 495 AudioCaps~\citep{kim2019audiocaps} validation clips. For each clip, we prompt ``Describe this audio'', record attention at every decoder layer, and use the LLM backbone to extract sound events aligned with generated tokens. Events are labeled real or hallucinated against the reference descriptions, and attention is averaged over each event's full core token set.

\textbf{Layer-wise attention allocation.} We first examine how attention is distributed among system, audio, and text tokens across decoder layers. As shown in Figure~\ref{fig:audio-attention-analysis}(a), audio attention exhibits a cluster of peaks in layers 15--22 of Audio Flamingo 3, indicating sustained attention to audio across this range. This motivates treating these peaks and middle layers as a contiguous window, termed the \emph{Audio Attention Stage} ($\mathcal{L}=\{15,\ldots,22\}$), for the subsequent In-Depth inspection. Appendix~\ref{app:audio_attention_stage} further explains the rationale for such windows and presents the corresponding Audio Attention Stages for other models.

\textbf{Attention mass.} We then compare attention patterns between real and hallucinated events. Both groups exhibit similar trends across layers: audio and text receive comparable attention, with their relative dominance alternating across layers. Averaged over the Audio Attention Stage, the attention mass assigned to audio is also comparable between the two groups (Figure~\ref{fig:audio-attention-analysis}(b)). Thus, hallucinated events can receive substantial audio attention.

\textbf{Attended positions.} Total attention mass does not specify where the model attends. Event-level attention maps (Figure~\ref{fig:audio-attention-analysis}(c)) show that real and hallucinated events with comparable audio attention can concentrate on different audio positions. This observation raises a further question: what do these differences in attended positions mean? Their significance depends on whether the attended representations semantically support the generated event.

\vspace{-5pt}

\begin{keyfinding}
 Real and hallucinated events receive comparable audio attention mass, yet their attention can concentrate on different audio positions. Substantial audio attention alone therefore does not distinguish real events from hallucinated ones.
\end{keyfinding}


\subsection{Semantic Mismatch Between Attended Audio Representations and Generated Events}

\begin{wrapfigure}{r}{0.48\textwidth}
    \vspace{-0.5\baselineskip} 
    \centering
    \includegraphics[width=\linewidth]{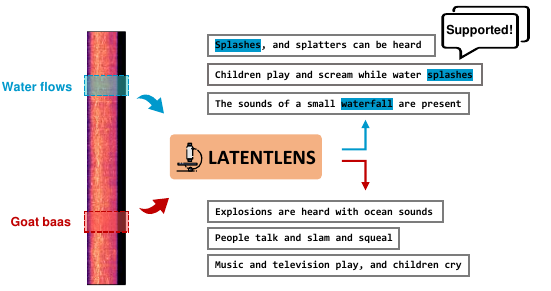}
    \captionsetup{
        font=small, 
        justification=raggedright, 
        singlelinecheck=false, 
        skip=4pt, 
        belowskip=-10pt 
    }
    \caption{Through LatentLens, the semantics decoded from high attention frames in the \emph{Audio Attention Stage} lend semantic support to real sound event tokens while withholding it from hallucinated sound event tokens.}
    \label{fig:semantic-consistency}
    \vspace{-0.5\baselineskip} 
\end{wrapfigure}

To assess whether attended audio positions semantically support generated events, we examine their audio representations. Vocabulary-based lens methods read hidden states through the output vocabulary: Logit Lens~\citep{nostalgebraist2020logitlens} uses the output projection directly, Tuned Lens~\citep{belrose2023eliciting} a learned transform. At audio positions, however, hidden states encode acoustic information and differ from text representations used for next-token prediction. These projections often produce incoherent or unrelated text—even Chinese tokens in English contexts—making semantic assessment difficult. We thus adapt LatentLens to audio: for each event, we retrieve contextualized text representations nearest to its highly attended audio representations and inspect their semantic support.

Figure~\ref{fig:semantic-consistency} shows retrieved contexts for the real event ``water flows'' and the hallucinated event ``goat baas.'' The contexts support the water event but suggest a different interpretation of the audio attended by the goat event. This contrast reveals a distinction between attention and semantic support: a model can attend to audio while generating an event inconsistent with the retrieved semantics of the attended representations.

\vspace{-5pt}
\begin{keyfinding}
Retrieved semantics match attended audio representations for real events, but mismatch them for hallucinated events.
\end{keyfinding}

\vspace{-5pt}
\subsection{Direct Audio Masking Can Remove Supported Content}

The previous subsection revealed a semantic mismatch between a hallucinated event and its attended audio representations. We next test how masking the corresponding audio regions affects both the hallucinated event and real events in the response.

\textbf{Simultaneous suppression of hallucinated and real events.} For the hallucinated event ``goat baas'' in Figure~\ref{fig:semantic-consistency}, we average attention across Audio-Attention layers, heads, and core tokens, select the ten most attended audio positions, and map them to approximate temporal windows. Zeroing the corresponding waveform samples and regenerating with the original prompt and decoding configuration yields ``A crowd of people are talking loudly with each other.'' The hallucinated goat event disappears, but the real water event also vanishes; only talking remains among the originally supported events. Local audio masking can thus suppress an unsupported event while removing supported content. A semantic mismatch with a hallucinated event does not imply that its attended audio is irrelevant to other events. Real and hallucinated events may attend to overlapping positions; masking solely from a hallucinated event's attention can disrupt evidence for real sounds. The intervention must therefore account for both regions to mask and regions to preserve. This motivates \emph{differential audio masking} in Sec.~\ref{sec:hallucination_mitigation}, where positions associated with predicted supported events are protected during suppression.

\begin{keyfinding}
Direct masking removes hallucinated events but also real ones. This trade-off motivates selecting masking regions while protecting those attended by real events.
\end{keyfinding}

\vspace{-5pt}
\section{Methods}
\vspace{-5pt}

\begin{figure*}[t]
    \centering
    \includegraphics[width=\textwidth]{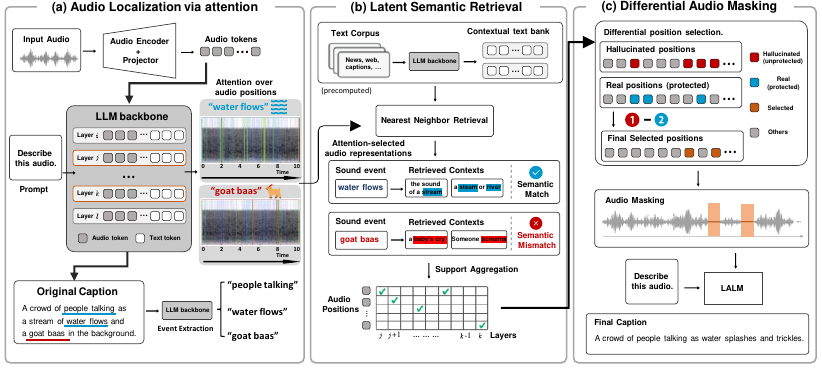}
    \caption{Overview of our detect-then-mitigate framework.
    \textbf{Detection (a, b):} event-level attention selects audio representations, and latent semantic retrieval assesses their support for generated events.
    \textbf{Mitigation (c):} Differential audio masking targets predicted hallucination regions, protects predicted supported-event regions, and regenerates the response.}
    \label{fig:method-workflow}
\end{figure*}

Motivated by the analysis in Sec.~\ref{sec:analysis}, we present a training-free framework for hallucination detection and mitigation (Figure~\ref{fig:method-workflow}). Detection combines attention-based audio localization with latent semantic retrieval to assess event-level support (a, b). Mitigation applies Differential Audio Masking before regenerating the response (c).

\vspace{-5pt}
\subsection{Hallucination Detection}
\label{subsec:hallucination_detection}
\vspace{-5pt}

Given audio $x$ and prompt $p$, a LALM $f_\theta$ generates a response $C=(c_1,\ldots,c_T)$. We retain its attention weights and audio hidden states to assess each generated event.

textbf{Event localization.}
\label{subsubsec:event_extraction}
\label{sec:audio_localization}
We prompt the model's own text LLM, without audio input, to extract event phrases and map them to ontology $\mathcal{O}$. Each event $e$ has a label $z_e$ and response-token positions $\mathcal{T}_e$ obtained by phrase alignment. Unmapped events remain unscored.
Let $A_{t,i}^{(\ell,h)}$ be the attention to audio position $i$ from head $h$ in layer $\ell$ at the step predicting $c_t$. Let $\mathcal{L}$ denote the layers in the Audio Attention Stage defined in Sec.~\ref{sec:analysis} (Figure~\ref{fig:audio-attention-analysis}(1)). We average over the $H$ heads and event tokens, then select the $k$ most attended positions in each layer $\ell\in\mathcal{L}$:
\begin{align}
    a_{e,i}^{(\ell)}
    &= \frac{1}{H|\mathcal{T}_e|}
       \sum_{h=1}^{H}\sum_{t\in\mathcal{T}_e} A_{t,i}^{(\ell,h)},
    \label{eq:event_attention}\\
    \mathcal{Q}_e^{(\ell)}
    &= \operatorname{TopK}_{i\in\{1,\ldots,N\}}
       \left(a_{e,i}^{(\ell)},k\right),
    \label{eq:audio_queries}
\end{align}
where $N$ is the number of audio tokens. The selected hidden states $h_i^{(\ell)}$ provide the queries for semantic retrieval.

\textbf{Latent semantic retrieval.}
\label{sec:latent_retrieval}
Following LatentLens~\citep{krojer2026latentlens}, we interpret audio states through neighboring contextualized text representations, avoiding direct vocabulary projection. We use the text captions from WavCaps~\citep{mei2024wavcaps} as the reference corpus. We build an offline bank $\mathcal{B}=\{(r_n^{(m)},d_n):m\in\mathcal{L}_{\mathcal{B}}\}$ by passing a text corpus through the frozen text LLM, where $r_n^{(m)}$ is a text-token state at layer $m$ and $d_n$ its context. The bank is reused across inputs. For each query, we retrieve $j$ neighbors across bank layers:
\begin{equation}
    \mathcal{N}_j(h_i^{(\ell)})
    = \underset{(r_n^{(m)},d_n)\in\mathcal{B}}{\operatorname{TopK}}
      \left(\phi_\ell(h_i^{(\ell)})^\top\phi_m(r_n^{(m)}),j\right).
    \label{eq:neighbor_retrieval}
\end{equation}
Here, $\phi_\ell$ applies preprocessing followed by $\ell_2$ normalization. Model-specific processing is detailed in Appendix~\ref{app:detection_configuration}.

\textbf{Support aggregation.}
\label{sec:support_aggregation}
The same text LLM maps retrieved contexts to ontology labels. Each selected layer--position pair contributes a binary vote $v_{e,i}^{(\ell)}$ if any retrieved label matches $z_e$ exactly or through an ancestor--descendant relation. We sum these votes and classify events below threshold $K$ as hallucinated:
\begin{align}
    S(e)
    &= \sum_{\ell\in\mathcal{L}}
       \sum_{i\in\mathcal{Q}_e^{(\ell)}}v_{e,i}^{(\ell)},
    \label{eq:event_support}\\
    \widehat{y}_e
    &= \mathbf{1}[S(e)<K].
    \label{eq:event_prediction}
\end{align}
The score satisfies $S(e)\in[0,|\mathcal{L}|k]$ regardless of how many neighbors match at each position. Retrieval depth $j$ controls the search, and $K$ sets the minimum required support. Appendix~\ref{app:support_aggregation} gives the formal matching and voting rules.

\vspace{-5pt}

\subsection{Hallucination Mitigation}
\label{sec:hallucination_mitigation}
\vspace{-5pt}

Events may attend to overlapping audio positions. \emph{Differential audio masking} protects positions associated with predicted supported events before selecting positions for intervention. We partition scored events into $\mathcal{E}_{\mathrm{hal}}=\{e:\widehat{y}_e=1\}$ and $\mathcal{E}_{\mathrm{sup}}=\{e:\widehat{y}_e=0\}$ using the detector's predictions.

\textbf{Differential position selection.}
\label{sec:event_protection}
\label{sec:differential_selection}
We average attention across selected layers and protect the $b_{\mathrm{p}}$ most attended positions of each event in $\mathcal{E}_{\mathrm{sup}}$:
\begin{align}
    \bar{a}_{e,i}
    &= \frac{1}{|\mathcal{L}|}\sum_{\ell\in\mathcal{L}}a_{e,i}^{(\ell)},
    \label{eq:mean_event_attention}\\
    \mathcal{P}
    &= \bigcup_{e\in\mathcal{E}_{\mathrm{sup}}}
       \operatorname{TopK}_{i}\left(\bar{a}_{e,i},b_{\mathrm{p}}\right).
    \label{eq:protected_positions}
\end{align}
For each predicted hallucination, we take the $b$ highest-ranked positions outside $\mathcal{P}$ and combine them:
\begin{align}
    \mathcal{M}_e
    &= \operatorname{TopK}_{i\in\{1,\ldots,N\}\setminus\mathcal{P}}
       \left(\bar{a}_{e,i},b\right),
    \label{eq:event_mask}\\
    \mathcal{M}
    &= \bigcup_{e\in\mathcal{E}_{\mathrm{hal}}}\mathcal{M}_e,
       \qquad \mathcal{M}\cap\mathcal{P}=\varnothing.
    \label{eq:joint_mask}
\end{align}
Here, $b_{\mathrm{p}}$ and $b$ are the protection and masking budgets, respectively. If $\mathcal{E}_{\mathrm{sup}}=\varnothing$, then $\mathcal{P}=\varnothing$.

\textbf{Audio Masking.}
\label{sec:caption_regeneration}
For an audio signal with $U$ samples, we map audio positions to uniform temporal bins, silence the selected bins, and regenerate with the original prompt and greedy decoding:
\begin{align}
    b_i
    &= \operatorname{round}\left(\frac{iU}{N}\right),
       \qquad i=0,\ldots,N,
    \label{eq:temporal_bins}\\
    \widetilde{x}[u]
    &= \begin{cases}
        0, & u\in\displaystyle\bigcup_{i\in\mathcal{M}}[b_{i-1},b_i),\\
        x[u], & \text{otherwise},
    \end{cases}
    \label{eq:waveform_intervention}\\
    \widetilde{C}
    &= \operatorname{GreedyDecode}(f_\theta,\widetilde{x},p).
    \label{eq:regeneration}
\end{align}
If $\mathcal{E}_{\mathrm{hal}}=\varnothing$, we retain the original response.

\section{Experiments}
\label{sec:experiments}

In this section, we empirically validate our framework. We first evaluate hallucination detection in Sec. \ref{sec:detection_experiments}, and then mitigation in Sec. \ref{sec:mitigation_experiments}.

\subsection{Hallucination Detection}
\label{sec:detection_experiments}
We first examine whether semantic support between the attended audio and generated content provides a reliable signal for detecting hallucinations. We evaluate this signal across datasets and models with different hallucination characteristics, and compare it against perplexity-, entropy-, attention-, and random-based baselines.
\subsubsection{Experimental Settings}

We evaluate hallucination detection on two benchmarks: the \texttt{eval} split of AudioSet Strong~\citep{Gemmeke2017} and the \texttt{fbsnr\_30dB} subset of DCASE2019 ~\citep{turpault2019sound, serizel2020sound}. We use three models: Audio Flamingo 3 (AF3)~\citep{ghosh2026audio}, Qwen2.5-Omni-7B~\citep{xu2025qwen25omnitechnicalreport}, and Step-Audio 2 mini~\citep{wu2025step}. For each dataset, we sample 200 clips for hyperparameter selection, shared across the three models. We then sample 500 clips from the remaining data for testing each dataset--model pair. We compare our detector against Perplexity, RAUQ Entropy, Attention Score, and Random. We report precision, recall, F1, and PR-AUC at both event and sample levels, treating hallucination as the positive class. Implementation details are provided in Appendix~\ref{app:detection_configuration}.

\begin{table*}[t]
\centering
\scriptsize
\setlength{\tabcolsep}{3.8pt}
\renewcommand{\arraystretch}{0.9}

\caption{
Hallucination detection results on AudioSet Strong and DCASE2019. Higher is better for Precision, Recall, F1, and PR-AUC. Best and second-best results within each model--level group are shown in bold and underlined, respectively.
}
\label{tab:detection_all}

\resizebox{\textwidth}{!}{
\begin{tabular}{@{}l|rrrrrr|rrrrrr@{}}
\toprule

\rowcolor{gray!15}
& \multicolumn{6}{c|}{\textbf{Event-Level}}
& \multicolumn{6}{c}{\textbf{Sample-Level}} \\

\rowcolor{gray!15}
\textbf{Method}
& \textbf{GT Hal.} & \textbf{Pred. Hal.}
& \textbf{Prec.} & \textbf{Recall} & \textbf{F1} & \textbf{PR-AUC}
& \textbf{GT Hal.} & \textbf{Pred. Hal.}
& \textbf{Prec.} & \textbf{Recall} & \textbf{F1} & \textbf{PR-AUC} \\
\midrule

\rowcolor{gray!25}
\multicolumn{13}{c}{\textbf{AudioSet Strong}} \\
\midrule

\multicolumn{13}{l}{\itshape Audio Flamingo 3} \\

Random
& \multirow{5}{*}{38.7} & 47.9
& 0.3241 & 0.4011 & 0.3585 & 0.3535
& \multirow{5}{*}{51.0} & 52.0
& 0.3846 & 0.3922 & 0.3883 & 0.4329 \\

Perplexity
& & 88.4
& 0.4307 & \textbf{0.9842}
& \underline{0.5992} & 0.5416
& & 94.4
& 0.5360 & \textbf{0.9922}
& 0.6960 & 0.6298 \\

RAUQ Entropy
& & 42.2
& \underline{0.5725} & 0.6253
& 0.5977 & \underline{0.5814}
& & 69.0
& \underline{0.6145} & 0.8314
& \underline{0.7067} & \underline{0.7037} \\

Attention Score
& & 58.0
& 0.4982 & \underline{0.7467}
& 0.5977 & 0.4733
& & 92.2
& 0.5445 & \underline{0.9843}
& 0.7011 & 0.5297 \\

\rowcolor{cyan!8}
\textbf{Ours}
& & 29.4
& \textbf{0.8646} & 0.6570
& \textbf{0.7466} & \textbf{0.7944}
& & 65.0
& \textbf{0.7046} & 0.8980
& \textbf{0.7897} & \textbf{0.9096} \\

\midrule

\multicolumn{13}{l}{\itshape Qwen2.5-Omni} \\

Random
& \multirow{5}{*}{31.1} & 48.2
& 0.3134 & 0.4865 & 0.3812 & 0.3236
& \multirow{5}{*}{50.2} & 49.0
& 0.4735 & 0.4622 & 0.4677 & 0.5029 \\

Perplexity
& & 93.5
& 0.3150 & \textbf{0.9485}
& 0.4729 & 0.3679
& & 100.0
& 0.5020 & \textbf{1.0000}
& 0.6684 & 0.5652 \\

RAUQ Entropy
& & 81.2
& 0.3292 & 0.8610
& 0.4763 & 0.4024
& & 99.0
& 0.5030 & \underline{0.9920}
& 0.6676 & \underline{0.5802} \\

Attention Score
& & 75.9
& \underline{0.3595} & \underline{0.8790}
& \underline{0.5103} & \underline{0.4407}
& & 95.0
& \underline{0.5116} & 0.9681
& \underline{0.6694} & 0.5101 \\

\rowcolor{cyan!8}
\textbf{Ours}
& & 32.3
& \textbf{0.5983} & 0.6229
& \textbf{0.6103} & \textbf{0.6067}
& & 61.4
& \textbf{0.6808} & 0.8327
& \textbf{0.7491} & \textbf{0.7537} \\

\midrule

\multicolumn{13}{l}{\itshape Step-Audio 2 mini} \\

Random
& \multirow{5}{*}{23.8} & 51.9
& 0.2510 & 0.5467 & 0.3441 & 0.2585
& \multirow{5}{*}{37.0} & 55.6
& 0.3273 & 0.4919 & 0.3931 & 0.3485 \\

Perplexity
& & 47.5
& 0.3415 & 0.6800
& 0.4547 & 0.3857
& & 77.4
& 0.4160 & \textbf{0.8703}
& 0.5629 & 0.5552 \\

RAUQ Entropy
& & 47.9
& \underline{0.3429} & 0.6889
& \underline{0.4579} & \underline{0.4110}
& & 64.2
& 0.4548 & 0.7892
& \underline{0.5771} & \textbf{0.6191} \\

Attention Score
& & 83.8
& 0.2668 & \textbf{0.9378}
& 0.4154 & 0.3535
& & 52.2
& \underline{0.4636} & 0.6541
& 0.5426 & 0.4922 \\

\rowcolor{cyan!8}
\textbf{Ours}
& & 40.1
& \textbf{0.4327} & \underline{0.7289}
& \textbf{0.5430} & \textbf{0.5010}
& & 55.8
& \textbf{0.5627} & \underline{0.8486}
& \textbf{0.6767} & \underline{0.6170} \\

\midrule
\rowcolor{gray!25}
\multicolumn{13}{c}{\textbf{DCASE2019}} \\
\midrule

\multicolumn{13}{l}{\itshape Audio Flamingo 3} \\

Random
& \multirow{5}{*}{2.7} & 49.1
& 0.0292 & \textbf{0.5263} & 0.0552 & 0.0743
& \multirow{5}{*}{3.2} & 56.6
& 0.0247 & \underline{0.4375} & 0.0468 & 0.0260 \\

Perplexity
& & 1.9
& 0.5385 & 0.3684 & 0.4375 & 0.3544
& & 1.4
& 0.7143 & 0.3125 & 0.4348 & 0.3393 \\

RAUQ Entropy
& & 1.6
& 0.6364 & 0.3684 & 0.4667 & 0.3726
& & 1.4
& 0.7143 & 0.3125 & 0.4348 & 0.3229 \\

Attention Score
& & 1.1
& \textbf{0.8750} & 0.3684
& \underline{0.5185} & \underline{0.3794}
& & 1.4
& \textbf{0.8571} & 0.3750
& \underline{0.5217} & \underline{0.3613} \\

\rowcolor{cyan!8}
\textbf{Ours}
& & 1.7
& \underline{0.7500} & \underline{0.4737}
& \textbf{0.5806} & \textbf{0.4761}
& & 2.2
& \underline{0.7273} & \textbf{0.5000}
& \textbf{0.5926} & \textbf{0.4806} \\

\midrule

\multicolumn{13}{l}{\itshape Qwen2.5-Omni} \\

Random
& \multirow{5}{*}{13.4} & 49.2
& 0.1459 & \underline{0.5340} & 0.2292 & 0.1415
& \multirow{5}{*}{12.8} & 54.4
& 0.1250 & 0.5313 & 0.2024 & 0.1693 \\

Perplexity
& & 31.3
& 0.2250 & 0.5243 & 0.3149 & \underline{0.2569}
& & 75.6
& 0.1455 & \textbf{0.8594}
& 0.2489 & 0.1652 \\

RAUQ Entropy
& & 70.0
& 0.1754 & \textbf{0.9126}
& 0.2942 & 0.2298
& & 68.6
& 0.1574 & \underline{0.8438}
& 0.2654 & 0.1498 \\

Attention Score
& & 23.1
& \underline{0.2655} & 0.4563
& \underline{0.3357} & 0.2086
& & 25.0
& \underline{0.2160} & 0.4219
& \underline{0.2857} & \underline{0.1726} \\

\rowcolor{cyan!8}
\textbf{Ours}
& & 14.0
& \textbf{0.4953} & 0.5146
& \textbf{0.5048} & \textbf{0.3337}
& & 14.2
& \textbf{0.5352} & 0.5938
& \textbf{0.5630} & \textbf{0.5422} \\

\midrule

\multicolumn{13}{l}{\itshape Step-Audio 2 mini} \\

Random
& \multirow{5}{*}{9.0} & 49.4
& 0.1104 & \textbf{0.6071} & 0.1868 & 0.1109
& \multirow{5}{*}{10.0} & 50.4
& 0.0952 & \underline{0.4800} & 0.1589 & 0.0979 \\

Perplexity
& & 10.6
& 0.4394 & \underline{0.5179}
& \underline{0.4754} & \underline{0.4387}
& & 7.0
& 0.4286 & 0.3000
& 0.3529 & 0.3253 \\

RAUQ Entropy
& & 5.6
& \underline{0.5714} & 0.3571
& 0.4396 & 0.4253
& & 30.6
& 0.2353 & \textbf{0.7200}
& 0.3547 & 0.3571 \\

Attention Score
& & 4.0
& \textbf{0.6400} & 0.2857
& 0.3951 & 0.3006
& & 7.2
& \underline{0.5278} & 0.3800
& \underline{0.4419} & \underline{0.3577} \\

\rowcolor{cyan!8}
\textbf{Ours}
& & 8.3
& 0.5577 & \underline{0.5179}
& \textbf{0.5370} & \textbf{0.5002}
& & 4.4
& \textbf{0.8182} & 0.3600
& \textbf{0.5000} & \textbf{0.5000} \\

\bottomrule
\end{tabular}
}

\end{table*}

\subsubsection{Detection Results and Analysis}

\textbf{Semantic support as a detection signal.} Table~\ref{tab:detection_all} shows that our detector achieves the highest F1 for all three models on both datasets, at both event and sample levels. Random provides a chance-level reference; our detector's consistently higher F1 and PR-AUC indicate discrimination beyond random guessing. PR-AUC also improves in 11 of the 12 comparisons, indicating better score ranking across thresholds. The exception is Step-Audio 2 mini at sample level on AudioSet Strong, where our PR-AUC is close to RAUQ Entropy's (0.6170 vs.\ 0.6191). These results support assessing whether attended audio representations semantically support the generated event as a useful alternative to attention magnitude and predictive uncertainty.

\textbf{The balance between precision and recall.} High recall can result from flagging most responses as hallucinated. For Qwen2.5-Omni on AudioSet Strong, Perplexity flags every response, although only 50.2\% contain hallucinations. Our detector raises precision from 0.5020 to 0.6808 while retaining 0.8327 recall. Thus, its higher F1 reflects a more selective identification of unsupported content, with fewer false alarms at the cost of some missed hallucinations. 

The tradeoff differs when hallucinations are rare. On DCASE2019, only 2.7\% of AF3 events are hallucinated. Attention Score achieves the highest precision but misses more hallucinations; our detector recovers more of them and achieves higher F1 at both levels. The gains therefore do not follow a uniform shift toward higher precision or recall. They reflect a better balance between the two under the evaluated settings, consistent with semantic support providing a useful signal for distinguishing real and hallucinated events.

\vspace{-5pt}
\subsection{Hallucination Mitigation}
\label{sec:mitigation_experiments}
Having established the effectiveness of semantic support for hallucination detection, we next investigate whether this signal can be used to mitigate hallucinations during generation. We compare differential audio masking with existing decoding- and attention-based approaches, evaluating both hallucination reduction and content preservation.
\vspace{-5pt}
\subsubsection{Experimental Settings}
\vspace{-5pt}

We evaluate mitigation on the same datasets and models as detection. For each dataset, we exclude the 200 hyperparameter-selection clips shared across models, then draw a fresh sample of 500 clips for each dataset--model pair from the remaining data to reduce reliance on a single draw for both detection and mitigation evaluation. We compare our method with Greedy decoding, Beam Search, Nucleus Sampling, AAD~\citep{hsu2025reducing}, and TCD~\citep{li2026temporal}. We report ECHO-I and ECHO-S for hallucination rates and Coverage for content retention. Implementation details are in Appendix~\ref{app:mitigation_configuration}.

\begin{table*}[t]
\centering
\scriptsize
\setlength{\tabcolsep}{6.0pt}
\renewcommand{\arraystretch}{0.92}

\caption{
Hallucination mitigation results on AudioSet Strong and DCASE2019. Metric values are percentages. Lower is better for ECHO-I and ECHO-S~\citep{kuan24_interspeech}, while higher is better for Coverage. Best and second-best results within each dataset--model group are shown in bold and underlined, respectively.
}
\label{tab:mitigation_results}

\begin{tabular}{@{}l|rrr|rrr@{}}
\toprule

\rowcolor{gray!15}
& \multicolumn{3}{c|}{\textbf{AudioSet Strong}}
& \multicolumn{3}{c}{\textbf{DCASE2019}} \\

\rowcolor{gray!15}
\textbf{Method}
& \textbf{ECHO-I} $\downarrow$
& \textbf{ECHO-S} $\downarrow$
& \textbf{Coverage} $\uparrow$
& \textbf{ECHO-I} $\downarrow$
& \textbf{ECHO-S} $\downarrow$
& \textbf{Coverage} $\uparrow$ \\
\midrule

\multicolumn{7}{l}{\itshape Audio Flamingo 3} \\

Greedy
& 22.7 & 34.0 & 31.8
& 2.9 & 3.4 & 60.1 \\

Beam
& 25.9 & 36.6 & 30.9
& 2.6 & \underline{2.8} & 60.0 \\

Nucleus
& 29.8 & 43.6 & 29.6
& 3.1 & 3.6 & 59.5 \\

AAD
& \underline{20.6} & \underline{29.2} & \textbf{34.2}
& \underline{2.5} & \underline{2.8} & \underline{62.3} \\

TCD
& 22.0 & 33.8 & 32.0
& 3.2 & 3.4 & \textbf{64.1} \\

\rowcolor{cyan!8}
\textbf{Ours}
& \textbf{14.9} & \textbf{23.0} & \underline{34.1}
& \textbf{2.2} & \textbf{2.6} & 60.3 \\

\midrule

\multicolumn{7}{l}{\itshape Qwen2.5-Omni} \\

Greedy
& 30.5 & 46.0 & 35.5
& 7.9 & 9.0 & 59.7 \\

Beam
& 32.3 & 51.2 & 33.3
& 9.6 & 11.6 & \textbf{62.0} \\

Nucleus
& 40.8 & 63.4 & 31.6
& 12.2 & 14.2 & 58.1 \\

AAD
& 28.9 & 45.2 & \underline{38.4}
& \underline{7.5} & \underline{7.8} & 55.6 \\

TCD
& \underline{28.5} & \underline{44.2} & 35.5
& 8.7 & 10.2 & \underline{61.7} \\

\rowcolor{cyan!8}
\textbf{Ours}
& \textbf{21.4} & \textbf{37.0} & \textbf{38.6}
& \textbf{6.5} & \textbf{6.6} & 54.7 \\

\midrule

\multicolumn{7}{l}{\itshape Step-Audio 2 mini} \\

Greedy
& \underline{19.7} & 31.6 & \textbf{30.5}
& 7.3 & 8.4 & \textbf{60.5} \\

Beam
& 20.6 & 30.2 & \underline{30.4}
& 6.9 & 8.0 & 59.7 \\

Nucleus
& 24.8 & 35.0 & 29.6
& \textbf{5.7} & \textbf{6.6} & 58.9 \\

AAD
& 20.6 & \underline{29.4} & 29.1
& \underline{6.4} & 7.6 & 58.5 \\

TCD
& 22.0 & 32.4 & 30.2
& 6.7 & 7.8 & \underline{59.8} \\

\rowcolor{cyan!8}
\textbf{Ours}
& \textbf{19.4}
& \textbf{28.6}
& 27.8
& \underline{6.4}
& \underline{7.4}
& 58.3\\

\bottomrule
\end{tabular}

\end{table*}

\subsubsection{Mitigation Results and Analysis}

\textbf{Targeted hallucination reduction.}
As shown in Table~\ref{tab:mitigation_results}, our method achieves the lowest ECHO-I and ECHO-S among standard decoding strategies and audio-aware baselines in five of the six dataset--model settings. On AudioSet Strong, it obtains the lowest hallucination rates across all three models. On DCASE2019, it leads on both metrics for AF3 and Qwen2.5-Omni. These results support the effectiveness of targeted audio masking across different models and datasets.

\textbf{Joint improvements in suppression and retention.}
Comparison across all methods shows that strong hallucination suppression can coexist with high content retention. On AudioSet Strong, our method achieves the best results on all three metrics for Qwen2.5-Omni. For AF3, it likewise achieves the lowest ECHO-I and ECHO-S, with Coverage just 0.1 percentage points below the highest value obtained by AAD (34.1\% vs.\ 34.2\%). These results show that the strongest hallucination reduction can be achieved with the highest or nearly highest Coverage among the compared methods. On DCASE2019, AF3 also achieves lower ECHO-I and ECHO-S than every baseline while retaining 60.3\% Coverage. This combination of suppression and retention is consistent with our protective design: differential masking suppresses attended audio positions while preserving those associated with predicted supported events. Ablations on the masking budget and selection strategy are provided in Appendices~\ref{app:masking_budget} and~\ref{app:masking_strategy}.

\vspace{-5pt}

\section{Conclusion}
\label{sec:conclusion}
In this work, we investigate sound event hallucination in LALMs through attention and latent representation senmantic analysis. We find that real and hallucinated events can receive comparable audio attention while attending to different positions. Their attended representations differ in their semantic match: retrieved textual contexts tend to align with grounded events but exhibit semantic mismatch with hallucinated ones. These observations motivate a training-free framework for hallucination detection and mitigation. Our detector identifies unsupported events by aggregating semantic support across audio positions and layers. Guided by these predictions, differential audio masking selectively intervenes on positions associated with hallucinated events while excluding those protected by supported events. Extensive experiments on AudioSet Strong and DCASE2019 with AF3, Qwen2.5-Omni and Step-Audio 2 mini demonstrate the effectiveness of this framework in detecting and reducing sound-event hallucinations. Our findings establish event-level semantic mismatch as a useful signal for understanding hallucination and guiding targeted intervention in LALMs.

\bibliography{iclr2027_conference}
\bibliographystyle{iclr2027_conference}

\newpage

\appendix

\section{Appendix}
This appendix provides method details and experimental settings, followed by additional validation and qualitative examples. Appendix~\ref{app:method_details} summarizes the full procedure and its components; Appendix~\ref{app:detection_configuration} specifies the evaluation protocol and implementation settings. Appendix~\ref{app:latentlens_reliability} evaluates the semantic validity of LatentLens retrieval. Appendices~\ref{app:detection_sensitivity} and~\ref{app:mitigation_ablation} examine sensitivity, masking choices, and response quality. Appendix~\ref{app:qualitative_cases} presents attention maps, retrieved contexts, and mitigation examples across the three models.

\subsection{Method Details}
\label{app:method_details}
We first summarize the complete detection and mitigation procedure, then detail the layer window used for semantic inspection and the ontology-based support votes.

\subsubsection{End-to-End Algorithm}
\label{app:end_to_end_algorithm}

Algorithm~\ref{alg:hallucination_detection_mitigation} summarizes the complete
detection and mitigation procedure described in Secs.~\ref{subsec:hallucination_detection}
and~\ref{sec:hallucination_mitigation}.

\SetKw{KwRet}{return}

\begin{algorithm}[h]
\caption{Semantic Support-Guided Detection and Differential Audio Masking.}
\label{alg:hallucination_detection_mitigation}
\KwIn{Audio $x$; prompt $p$; frozen LALM $f_\theta$; ontology $\mathcal O$;
text bank $\mathcal B$; selected layers $\mathcal L$; audio-position budget $k$;
retrieval depth $j$; support threshold $K$; protection budget $b_{\mathrm p}$;
masking budget $b$}
\KwOut{Response $C^*$; event predictions $\{\widehat y_e\}$; mask $\mathcal M$}

$(C,A,H)\leftarrow \operatorname{Generate}(f_\theta,x,p,
\text{return attention and hidden states})$\;
$\mathcal E\leftarrow \operatorname{ExtractMapAlign}(C,\mathcal O)$\tcp*{mapped events with token spans $\mathcal T_e$}

\ForEach{$e\in\mathcal E$}{
  $S(e)\leftarrow 0$\;
  \ForEach{$\ell\in\mathcal L$}{
    \For{$i\leftarrow 1$ \KwTo $N$}{
      $a_{e,i}^{(\ell)}\leftarrow
      \frac{1}{H_{\mathrm{heads}}|\mathcal T_e|}
      \sum_{h=1}^{H_{\mathrm{heads}}}\sum_{t\in\mathcal T_e}
      A_{t,i}^{(\ell,h)}$\;
    }
    $\mathcal Q_e^{(\ell)}\leftarrow
    \operatorname{TopK}_{i}(a_{e,i}^{(\ell)},k)$\;
    \ForEach{$i\in\mathcal Q_e^{(\ell)}$}{
      $\mathcal N\leftarrow
      \operatorname{Retrieve}_{j}(\phi_\ell(h_i^{(\ell)}),\mathcal B)$\;
      $\mathcal Z\leftarrow \operatorname{MapContexts}(\mathcal N,\mathcal O)$\;
      $v_{e,i}^{(\ell)}\leftarrow
      \mathbf 1[\exists z\in\mathcal Z:\operatorname{OntMatch}(z,z_e)=1]$\;
      $S(e)\leftarrow S(e)+v_{e,i}^{(\ell)}$\;
    }
  }
  $\widehat y_e\leftarrow \mathbf 1[S(e)<K]$\tcp*{$1$: hallucinated; $0$: supported}
}

$\mathcal E_{\mathrm{hal}}\leftarrow\{e:\widehat y_e=1\}$;
$\mathcal E_{\mathrm{sup}}\leftarrow\{e:\widehat y_e=0\}$\;
\If{$\mathcal E_{\mathrm{hal}}=\varnothing$}{
  \KwRet $(C,\{\widehat y_e\},\varnothing)$\;
}

\ForEach{$e\in\mathcal E$}{
  $\bar a_{e,i}\leftarrow |\mathcal L|^{-1}
  \sum_{\ell\in\mathcal L}a_{e,i}^{(\ell)}$ for all $i$\;
}
$\mathcal P\leftarrow
\bigcup_{e\in\mathcal E_{\mathrm{sup}}}
\operatorname{TopK}_{i}(\bar a_{e,i},b_{\mathrm p})$\;
$\mathcal M\leftarrow\varnothing$\;
\ForEach{$e\in\mathcal E_{\mathrm{hal}}$}{
  $\mathcal M_e\leftarrow
  \operatorname{TopK}_{i\in\{1,\ldots,N\}\setminus\mathcal P}
  (\bar a_{e,i},b)$\tcp*{skip protected positions and continue until $b$ positions are selected}
  $\mathcal M\leftarrow\mathcal M\cup\mathcal M_e$\;
}
$\widetilde x\leftarrow
\operatorname{SilenceBins}(x,\mathcal M,N)$\tcp*{mask only the central half of each bin for Step-Audio 2 mini}
$\widetilde C\leftarrow
\operatorname{GreedyDecode}(f_\theta,\widetilde x,p)$\;
\KwRet $(\widetilde C,\{\widehat y_e\},\mathcal M)$\;
\end{algorithm}

\subsubsection{Audio Attention Stage}
\label{app:audio_attention_stage}
\begin{figure*}[htbp]
\centering
\begin{minipage}[t]{0.48\textwidth}
\centering
\includegraphics[width=\linewidth]{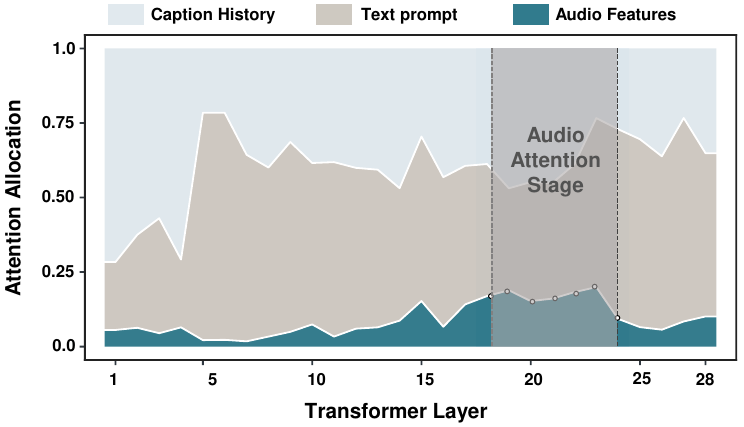}
\par\smallskip
{\small (a) Qwen2.5-Omni-7B}
\end{minipage}\hfill
\begin{minipage}[t]{0.48\textwidth}
\centering
\includegraphics[width=\linewidth]{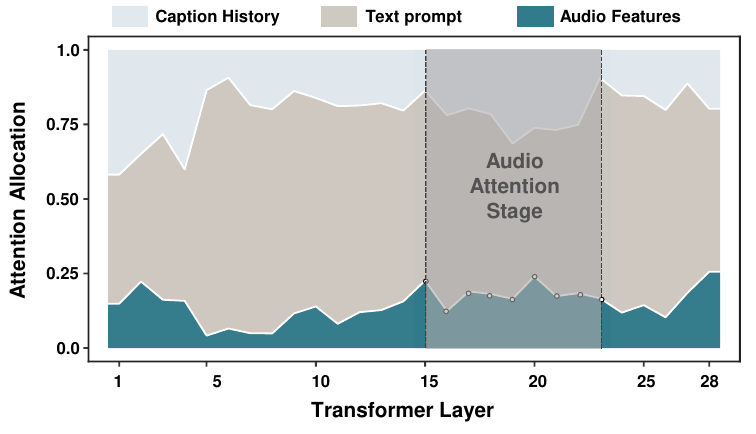}
\par\smallskip
{\small (b) Step-Audio 2 mini}
\end{minipage}
\caption{Layer-wise attention allocation on 495 AudioCaps validation clips
for each model. Attention to audio features, non-audio prompt tokens, and
caption history is averaged over heads and caption-generation decisions
within each clip, then equally across clips. Shaded intervals indicate the
Audio Attention Stage: L18--24 for Qwen2.5-Omni-7B and L15--23 for
Step-Audio 2 mini. The profiles include all caption-generation decisions,
not just tokens belonging to sound events.}
\label{fig:appendix_audio_attention_stage}
\end{figure*}

We use the Audio Attention Stage as a model-specific window for inspecting
the audio representations attended to during caption generation. The layer
ranges are 15--22 for Audio Flamingo 3, 18--24 for Qwen2.5-Omni-7B, and
15--23 for Step-Audio 2 mini. All indices are 1-based and refer to the text
decoder, which has 28 layers in each of these checkpoints.
Figure~\ref{fig:appendix_audio_attention_stage} complements the Audio Flamingo 3
profile in Figure~\ref{fig:audio-attention-analysis}(a).

\textbf{Layer-wise motivation.}
For Qwen2.5-Omni-7B, audio receives a small share of attention in the early
layers, followed by a more sustained increase at L18--23. The decrease at
L24 marks the transition toward the lower audio allocation observed at
L25--26. The L18--24 window covers this elevated region and its transition.
Step-Audio 2 mini has a less uniform profile: the L15--23 window includes
peaks at L15 and L20 together with the intervening layers, followed by lower
audio attention at L24--26. Audio attention rises again at L27--28, so the
selected window does not encompass every layer with high audio attention.
These profiles motivate inspecting a contiguous group of layers with
repeated access to audio, instead of interpreting an isolated attention peak.

\textbf{Role in semantic inspection.}
Within each selected layer, we average attention over heads and the core
tokens of an event, identify its most-attended audio positions, and inspect
their hidden states through Audio LatentLens. A contiguous window permits
support evidence to be aggregated across several representations of the
same event. The stage is defined relative to each model's layer-wise profile;
audio need not receive more attention than prompt tokens or caption history.
Attention mass identifies positions to inspect, while the retrieved semantics
determine whether those representations support the generated event.

\subsubsection{Semantic Support Aggregation}
\label{app:support_aggregation}
For each retrieved context $d$, the model's text LLM extracts sound events and maps them to ontology labels $\mathcal{Z}(d)$. Each distinct context is annotated once and cached. To accommodate different levels of specificity, labels match if they are identical or one is an ancestor of the other:
\begin{equation}
    \rho(z,z')=\mathbf{1}\left[
        z=z' \;\lor\; z\in\operatorname{Anc}(z')
        \;\lor\; z'\in\operatorname{Anc}(z)
    \right],
    \label{eq:ontology_matching}
\end{equation}
where $\operatorname{Anc}(z)$ contains the ancestors of $z$. Siblings do not match through a shared ancestor, and a flat vocabulary uses exact matching.
Each selected layer--position pair casts one support vote if any retrieved context contains a compatible label:
\begin{equation}
    v_{e,i}^{(\ell)}=\mathbf{1}\left[
        \exists\,(r,d)\in\mathcal{N}_j(h_i^{(\ell)}),
        \ \exists\,z'\in\mathcal{Z}(d):\rho(z_e,z')=1
    \right].
    \label{eq:support_vote}
\end{equation}

Each layer--position pair contributes at most one vote, regardless of how many retrieved contexts match. The support score and decision rule are given in Eqs.~\ref{eq:event_support}--\ref{eq:event_prediction}. Thus, $S(e)\in[0,|\mathcal{L}|k]$, and $K$ specifies the minimum required support.

\subsection{Experimental Setup and Implementation Details}
\label{app:detection_configuration}

\textbf{Data and response generation.}
We use the \texttt{eval} split of AudioSet Strong with its hierarchical
sound-event vocabulary and the \texttt{fbsnr\_30dB} split of DCASE2019.
The latter contains ten-second domestic sound mixtures with ten reference
classes. For Audio Flamingo 3, Qwen2.5-Omni-7B, and Step-Audio 2 mini, we
generate responses with each model's audio-description prompt, extract
sound events, and map them to the corresponding dataset vocabulary.

\textbf{Ground-truth labels.}
We label a mapped event as hallucinated if no reference label supports
it. AudioSet Strong uses ancestor-or-descendant matching; DCASE2019 uses
exact matching. We apply these rules to both detection and mitigation. We
exclude unmapped events from event-level scoring. We label a response
positive at the sample level if it contains at least one unsupported
mapped event.

\textbf{Evaluation splits.}
For each dataset, we sample 200 clips from the designated split for
detector hyperparameter selection and freeze the selected parameters. We
exclude these clips from the evaluation pool and then sample 500 clips
from the remaining data for detection testing for each dataset--model
pair. We also exclude the 200 selection clips from mitigation evaluation.
We return the detection clips to the evaluation pool before drawing 500
mitigation clips; the two evaluation sets may therefore overlap. Within
each experiment, all methods use the same clips.

\textbf{Detection evaluation.}
We compare against Perplexity, RAUQ Entropy, Attention Score, and Random
at event and sample levels. Hallucination is the positive class for
precision, recall, F1, and PR-AUC. We compute F1 at the frozen decision
threshold; PR-AUC measures score ranking across thresholds. In
Table~\ref{tab:detection_all}, GT Hal.\ and Pred.\ Hal.\ denote the
reference and predicted positive rates, reported as percentages.

\textbf{Detector settings.}
Table~\ref{tab:detection_parameters} lists the event-level settings for
each dataset and model. We use layers within the model's Audio-Attention
Stage (Appendix~\ref{app:audio_attention_stage}) and fix $k=3$ attended
audio positions per layer. The retrieval budget $j$ and support threshold
$K$ vary by dataset and model, as shown in the table. All main experiments use cross-layer retrieval.
Audio Flamingo 3 and Step-Audio 2 mini use direct normalization. Qwen2.5-Omni-7B first subtracts the text-bank mean at the representation's layer.
\begin{table*}[h]
\centering
\begin{minipage}[t]{0.57\textwidth}
\vspace{0pt}
\centering
\small
\setlength{\tabcolsep}{3pt}
\begin{tabular}{@{}llccc@{}}
\toprule
Dataset & Model & $k$ & $j$ & $K$ \\
\midrule
\multirow{3}{*}{AudioSet Strong}
  & Audio Flamingo 3  & 3 & 3 & 5  \\
  & Qwen2.5-Omni      & 3 & 3 & 4  \\
  & Step-Audio 2 mini & 3 & 4 & 13 \\
\midrule
\multirow{3}{*}{DCASE2019}
  & Audio Flamingo 3  & 3 & 5 & 4 \\
  & Qwen2.5-Omni      & 3 & 4 & 9 \\
  & Step-Audio 2 mini & 3 & 4 & 6 \\
\bottomrule
\end{tabular}
\caption{Event-level detection configurations.}
\label{tab:detection_parameters}
\end{minipage}
\hfill
\begin{minipage}[t]{0.40\textwidth}
\vspace{0pt}
$k$ is the number of attended audio positions selected per layer.
$j$ is the number of retrieved text neighbors per position.
$K$ is the minimum number of semantic-support votes required to
accept an event. Each selected layer--position pair contributes
at most one vote, as defined in Appendix~\ref{app:support_aggregation}.
\end{minipage}
\end{table*}

\textbf{Mitigation settings.}\label{app:mitigation_configuration} We keep the event-level detector configurations in Table~\ref{tab:detection_parameters} fixed throughout mitigation. For both datasets and all three models, we use top10/min10: we protect the ten most attended positions of each event predicted to be reliable, then select ten unprotected positions for each predicted hallucinated event. When a protected position is skipped, we continue down the event's attention ranking until ten positions are selected. We silence the approximate audio windows corresponding to their union and regenerate with greedy decoding and the original prompt. Step-Audio 2 mini has approximately twice the temporal bin width of AF3 and Qwen2.5-Omni. For Step, we therefore halve each mapped masking window symmetrically about its center, masking only the central half of the bin. If no hallucination is detected, we retain the original response. 

\textbf{Mitigation evaluation.} We map events in the resulting responses to the dataset vocabulary and score them against the original audio's reference labels, using the matching rules above. Each mapped class is counted once per response; unmapped descriptions are excluded from class-level counts. Following the ECHO and Cover definitions in~\citet{kuan24_interspeech}, we compute corpus-level percentages using the dataset-specific matching rules. For response $n$, let $\mathcal{P}_n$ be its set of mapped predicted classes and $\mathcal{G}_n$ the original audio's reference classes. Let $\mathcal{H}_n\subseteq\mathcal{P}_n$ contain predictions that match no reference class, and let $\mathcal{M}_n$ be a maximum one-to-one matching between $\mathcal{P}_n$ and $\mathcal{G}_n$ under the same matching rules. Over $N_{\mathrm{eval}}$ responses, the metrics are
\begin{align}
    \mathrm{ECHO\text{-}I}
    &= 100\,\frac{\sum_{n=1}^{N_{\mathrm{eval}}}|\mathcal{H}_n|}
    {\sum_{n=1}^{N_{\mathrm{eval}}}|\mathcal{P}_n|},
    \label{eq:echo_i}\\
    \mathrm{ECHO\text{-}S}
    &= \frac{100}{N_{\mathrm{eval}}}
    \sum_{n=1}^{N_{\mathrm{eval}}}\mathbf{1}[|\mathcal{H}_n|>0],
    \label{eq:echo_s}\\
    \mathrm{Coverage}
    &= 100\,\frac{\sum_{n=1}^{N_{\mathrm{eval}}}|\mathcal{M}_n|}
    {\sum_{n=1}^{N_{\mathrm{eval}}}|\mathcal{G}_n|}.
    \label{eq:mitigation_coverage}
\end{align}
ECHO-I measures the fraction of predicted classes that are unsupported; ECHO-S measures the fraction of responses containing any unsupported class. Coverage measures the fraction of reference classes covered by the matched predictions. One-to-one matching prevents multiple related predictions from counting the same reference class more than once. We report hallucination rates together with Coverage to assess whether suppression preserves supported content.

\subsection{Reliability of LatentLens for Audio}
\label{app:latentlens_reliability}
We evaluate the reliability of mapping audio latent representations to textual explanations through LatentLens. Table~\ref{tab:semantic_validity} reports semantic validity on AudioTime~\citep{xie2025audiotime} test set using ancestor-or-descendant matching rules on  Audioset ontology. For each model, we select layers within its Audio Attention Stage. Same-layer retrieval searches text vectors from the audio query's decoder layer, whereas cross-layer retrieval searches the union of text vectors from the selected layers. Both settings are evaluated with real audio, silence, random retrieval, and shuffled ground truth.

Both clip-level and token-level evaluations use Uniform-8 audio-token sampling. Token neighbors are retained without text deduplication, and clip candidates are pooled without text or vector deduplication. Random retrieval follows the predefined token-row sampling control.

All scores are percentages. Clip-level Hit@$K$, for $K\in\{1,5,10\}$, averages layer-wise class-macro scores. For token-level Interpretability@5, let $\mathcal{N}_5(i,t,\ell)$ denote the top-five text neighbors of token $t$ from clip $i$ at layer $\ell$, and let $E(n)$ denote the event labels annotated for neighbor $n$. A token--layer pair is counted as interpretable when at least one retrieved event matches the clip label $y_i$:
\begin{equation}
z_{i,t,\ell}=\mathbf{1}\!\left[
\exists n\in\mathcal{N}_5(i,t,\ell),\
e\in E(n):
\operatorname{match}(e,y_i)=1
\right],
\end{equation}
where $\operatorname{match}$ accepts identical labels or an ancestor--descendant relation. Micro scores pool all evaluated token--layer pairs $\mathcal{Q}$:
\begin{equation}
I_{\mathrm{micro}}@5
=
\frac{100}{|\mathcal{Q}|}
\sum_{(i,t,\ell)\in\mathcal{Q}} z_{i,t,\ell}.
\end{equation}
Macro scores average token hits within each class, then equally across classes and layers:
\begin{equation}
I_{\mathrm{macro}}@5
=
\frac{100}{|\mathcal{L}|}
\sum_{\ell\in\mathcal{L}}
\frac{1}{|\mathcal{C}_{\ell}|}
\sum_{c\in\mathcal{C}_{\ell}}
\frac{\sum_{(i,t)\in\mathcal{Q}_{\ell,c}}z_{i,t,\ell}}
{|\mathcal{Q}_{\ell,c}|},
\end{equation}
where $\mathcal{Q}_{\ell,c}$ contains the evaluated tokens from clips labeled $c$ at layer $\ell$, and $\mathcal{C}_{\ell}$ is the set of evaluated classes. Token-level scores measure agreement with clip-level labels.

\begin{table*}[h]
\centering
\caption{Semantic validity on AudioTime with ancestor-or-descendant matching.
AF3 = Audio Flamingo 3; Qwen = Qwen2.5-Omni-7B; Step = Step-Audio 2 Mini.
Real = real audio; Sil. = silence; Rand. = random retrieval; Shuf. = shuffled ground truth.
}
\label{tab:semantic_validity}
\small
\setlength{\tabcolsep}{5.0pt}
\renewcommand{\arraystretch}{1.12}
\begin{tabular}{lllrrrrr}
\toprule
\multirow{2}{*}{Model} & \multirow{2}{*}{Retrieval} & \multirow{2}{*}{Cond.}
& \multicolumn{3}{c}{Clip-level Hit@K}
& \multicolumn{2}{c}{Token-level Interpretability@5} \\
\cmidrule(lr){4-6}\cmidrule(lr){7-8}
& & & H@1 & H@5 & H@10 & Micro & Macro \\
\midrule
\multirow{8}{*}{Audio Flamingo 3} & \multirow{4}{*}{Same-layer} & Real & 49.27 & 66.96 & 73.43 & 58.17 & 56.90 \\
 &  & Sil. & 0.89 & 5.96 & 10.84 & 4.69 & 4.86 \\
 &  & Rand. & 1.21 & 5.98 & 11.24 & 5.65 & 6.00 \\
 &  & Shuf. & 2.60 & 7.59 & 11.99 & 7.74 & 7.62 \\
\cmidrule(lr){2-8}
 & \multirow{4}{*}{Cross-layer} & Real & 49.46 & 68.03 & 77.78 & 64.41 & 63.14 \\
 &  & Sil. & 1.19 & 4.20 & 7.09 & 2.97 & 2.91 \\
 &  & Rand. & 1.21 & 5.98 & 11.24 & 5.65 & 6.00 \\
 &  & Shuf. & 2.79 & 5.98 & 8.75 & 5.46 & 5.42 \\
\midrule
\multirow{8}{*}{Qwen2.5-Omni-7B} & \multirow{4}{*}{Same-layer} & Real & 44.59 & 60.89 & 68.88 & 57.19 & 56.93 \\
 &  & Sil. & 0.41 & 2.56 & 6.42 & 3.72 & 3.93 \\
 &  & Rand. & 1.17 & 5.99 & 10.93 & 5.66 & 5.80 \\
 &  & Shuf. & 2.05 & 5.91 & 9.83 & 6.16 & 6.20 \\
\cmidrule(lr){2-8}
 & \multirow{4}{*}{Cross-layer} & Real & 44.65 & 62.11 & 72.20 & 59.01 & 58.75 \\
 &  & Sil. & 0.34 & 1.71 & 3.09 & 2.69 & 2.86 \\
 &  & Rand. & 1.17 & 5.99 & 10.93 & 5.66 & 5.80 \\
 &  & Shuf. & 1.91 & 4.60 & 7.20 & 4.53 & 4.56 \\
\midrule
\multirow{8}{*}{Step-Audio 2 mini} & \multirow{4}{*}{Same-layer} & Real & 35.83 & 54.87 & 61.77 & 54.74 & 52.35 \\
 &  & Sil. & 1.74 & 9.41 & 15.49 & 8.60 & 7.38 \\
 &  & Rand. & 1.25 & 5.54 & 10.18 & 5.52 & 5.72 \\
 &  & Shuf. & 1.48 & 6.40 & 12.18 & 8.65 & 7.48 \\
\cmidrule(lr){2-8}
 & \multirow{4}{*}{Cross-layer} & Real & 36.03 & 55.22 & 63.42 & 58.46 & 56.23 \\
 &  & Sil. & 1.67 & 7.52 & 11.33 & 5.62 & 4.84 \\
 &  & Rand. & 1.25 & 5.54 & 10.18 & 5.52 & 5.72 \\
 &  & Shuf. & 1.34 & 4.37 & 6.33 & 5.95 & 5.13 \\
\bottomrule
\end{tabular}
\end{table*}

\subsection{Detection Sensitivity Analysis}
\label{app:detection_sensitivity}
We assess local parameter perturbations, retrieval budgets, and ontology matching rules. The ontology analysis also reports their effect on mitigation evaluation.

\subsubsection{Precision--Recall Curves for Hallucination Detection}
\label{app:precision_recall_curves}
\begin{figure*}[t]
\centering
\includegraphics[width=\textwidth]{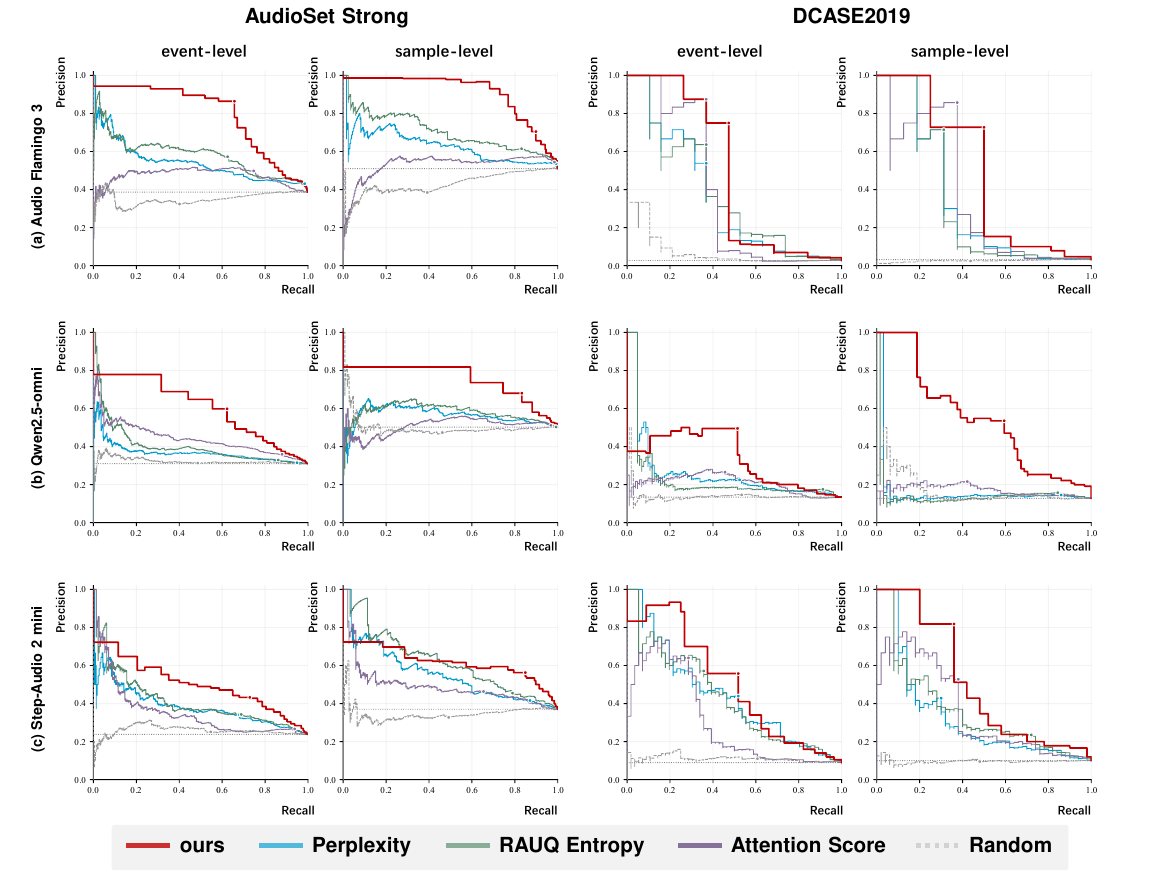}
\caption{Precision--recall curves for hallucination detection on AudioSet Strong (left column pair) and DCASE2019 (right column pair). Rows correspond to Audio Flamingo 3, Qwen2.5-Omni, and Step-Audio 2 mini. Within each dataset, event-level and sample-level results appear on the left and right, respectively. Colors identify the five methods; markers indicate the operating points in Table~\ref{tab:detection_all}. Dotted horizontal lines indicate positive-class prevalence. Each panel uses the same evaluation items across methods, on the model-specific post-hoc selected cohorts underlying the table.}
\label{fig:appendix_detection_pr}
\end{figure*}

Figure~\ref{fig:appendix_detection_pr} shows the precision--recall curves corresponding to the AudioSet Strong and DCASE2019 results for all three models in Table~\ref{tab:detection_all}, at both event and sample levels. We use archived precision--recall points for Audio Flamingo 3 and Qwen2.5-Omni on AudioSet Strong. For Step-Audio 2 mini on AudioSet Strong and all three models on DCASE2019, we recover scores from cached features and semantic retrieval results on the corresponding 500 sample IDs; each recovered score configuration reproduces the table's precision, recall, F1, and AP to four decimal places. The compatible baseline score variants are unique within the existing implementation. We keep each score configuration fixed and include tied scores together when sweeping thresholds. Hallucination is the positive class. The quantity reported as PR-AUC in the table is non-interpolated average precision (AP), computed as $\sum_i (R_i-R_{i-1})P_i$, rather than trapezoidal area. We plot the empirical step curves without smoothing; markers identify the operating points reported in the table.

\textbf{Ranking quality and operating points.}
On AudioSet Strong, our AP for Audio Flamingo 3 is 0.7944 at event level and 0.9096 at sample level, compared with 0.5814 and 0.7037 for the strongest baseline, RAUQ Entropy. For Qwen2.5-Omni, our AP is 0.6067 and 0.7537, compared with the strongest baseline values of 0.4407 (Attention Score) and 0.5802 (RAUQ Entropy), respectively. These differences quantify ranking performance across thresholds; they do not imply that one curve has higher precision at every recall. At the reported sample-level operating point for Qwen2.5-Omni, Perplexity achieves recall 1.0000 with precision 0.5020, whereas our detector achieves recall 0.8327 with precision 0.6808. Thus, the selected threshold trades some recall for fewer false positives.

\textbf{Interpretation and scope.}
The semantic-support score is discrete, so sweeping its threshold produces fewer distinct operating points than continuous uncertainty scores. Its larger steps should therefore be interpreted as tied groups entering the prediction set together. Positive-class prevalence also differs across panels, making within-panel comparisons the appropriate reference. The source records identify these cohorts as selected after inspecting outcomes; the curves describe the corresponding table results and are not an additional independent test. Figure~\ref{fig:appendix_detection_pr} covers all twelve dataset--model--level settings. Step-Audio 2 mini illustrates the distinction between ranking quality and thresholded decisions: our event-level AP is 0.5010, compared with 0.4110 for RAUQ Entropy, whereas our sample-level AP is slightly lower (0.6170 versus 0.6191) despite higher F1 at the reported operating point (0.6767 versus 0.5771). Thus, the sample-level F1 gain should not be interpreted as uniformly better ranking across thresholds. On DCASE2019, our detector achieves the highest AP in all six settings: 0.4761/0.4806 for Audio Flamingo 3, 0.3337/0.5422 for Qwen2.5-Omni, and 0.5002/0.5000 for Step-Audio 2 mini (event/sample). The low positive prevalence, particularly for Audio Flamingo 3, produces larger recall increments when individual positive examples enter the prediction set; the jagged empirical curves should therefore be read together with AP and the marked operating points.

\subsubsection{Local Parameter Sensitivity}
\label{app:audioset_detection_sensitivity}
We perturb one parameter at a time around the respective event- and sample-level reference configurations, keeping sample IDs, responses, ground-truth labels, and all other parameters fixed. We vary $K$ and $j$ by one, remove either boundary layer, and shift the layer window by one.

Table~\ref{tab:audioset_detection_sensitivity} reports event- and sample-level F1 ranges on the 500-clip detection evaluation set for each model. Perturbing $j$ or $K$ changes F1 by $-4.50$ to $+0.75$ percentage points relative to the reference configuration. 

\begin{table*}[h]
\centering\small
\caption{Local sensitivity of hallucination detection on AudioSet Strong. F1 is reported on a 0--1 scale. Ranges contain all available perturbed configurations for that factor, excluding the reference. Layer perturbations remove either boundary layer or shift the window by one within cached coverage. All unperturbed parameters remain fixed.}
\label{tab:audioset_detection_sensitivity}
\resizebox{\textwidth}{!}{%
\begin{tabular}{llccccc}
\toprule
Model & Level & Reference F1 & $K\pm1$ & $j\pm1$ & Layer perturbations & $k=2$ \\
\midrule
\multirow{2}{*}{Audio Flamingo 3} & Event & 0.7466 & 0.7016--0.7206 & 0.7242--0.7244 & 0.7296--0.7352 & 0.6930 \\
& Sample & 0.7897 & 0.7667--0.7972 & 0.7695-0.7695 & 0.7568--0.7727 & 0.7168 \\
\multirow{2}{*}{Qwen2.5-Omni} & Event & 0.6103 & 0.5880--0.5978 & 0.5954--0.6096 & 0.5885--0.5961 & 0.5717 \\
 & Sample & 0.7491 & 0.7367--0.7406 & 0.7312--0.7434 & 0.7159--0.7435 & 0.7227 \\
\multirow{2}{*}{Step-Audio 2 mini} & Event & 0.5430 & 0.5359--0.5486 & 0.5359--0.5440 & 0.5329--0.5412 & 0.4844 \\
& Sample & 0.6767 & 0.6639--0.6682 & 0.6626--0.6728 & 0.6612--0.6680 & 0.6333 \\
\bottomrule
\end{tabular}}
\end{table*}

\subsubsection{Sensitivity to Retrieval Budgets}
\label{app:retrieval_sensitivity}

We vary the audio-token budget $k \in \{1,2,3\}$
and the text-neighbor budget $j \in \{1,2,3,4,5\}$,
while keeping the selected layer range fixed for each model.
We report F1 and PR-AUC. Results on AudioSet Strong are listed in Table ~\ref{tab:kj_sensitivity_audioset}.

\begin{table*}[h]
\centering
\caption{Sensitivity on AudioSet Strong to audio-token budget $k\in\{1,2,3\}$ and text-neighbor budget $j\in\{1,2,3,4,5\}$. The layer range is fixed. Gray cells mark the $(k,j)$ choices achieve the results. Higher values are better.}
\label{tab:kj_sensitivity_audioset}
\scriptsize
\setlength{\tabcolsep}{3.5pt}
\renewcommand{\arraystretch}{1.12}
\begin{tabular}{@{}cc *{12}{c}@{}}
\toprule
& & \multicolumn{4}{c}{Audio Flamingo 3} & \multicolumn{4}{c}{Qwen2.5-Omni-7B} & \multicolumn{4}{c}{Step-Audio 2 mini} \\
\cmidrule(lr){3-6}\cmidrule(lr){7-10}\cmidrule(lr){11-14}
& & \multicolumn{2}{c}{Event-level} & \multicolumn{2}{c}{Sample-level} & \multicolumn{2}{c}{Event-level} & \multicolumn{2}{c}{Sample-level} & \multicolumn{2}{c}{Event-level} & \multicolumn{2}{c}{Sample-level} \\
\cmidrule(lr){3-4}\cmidrule(lr){5-6}\cmidrule(lr){7-8}\cmidrule(lr){9-10}\cmidrule(lr){11-12}\cmidrule(lr){13-14}
$k$ & $j$ & F1 & PR-AUC & F1 & PR-AUC & F1 & PR-AUC & F1 & PR-AUC & F1 & PR-AUC & F1 & PR-AUC \\
\midrule
\multirow{5}{*}{1} & 1 & 0.6058 & 0.6185 & 0.6755 & 0.7385 & 0.5151 & 0.4643 & 0.6849 & 0.5782 & 0.3849 & 0.3845 & 0.5449 & 0.5389 \\
 & 2 & 0.6242 & 0.6641 & 0.6755 & 0.7922 & 0.5246 & 0.4910 & 0.6906 & 0.6026 & 0.3849 & 0.4062 & 0.5449 & 0.5639 \\
 & 3 & 0.6388 & 0.6881 & 0.6755 & 0.8195 & 0.5341 & 0.5138 & 0.6910 & 0.6178 & 0.3849 & 0.4196 & 0.5449 & 0.5724 \\
 & 4 & 0.6388 & 0.6950 & 0.6755 & 0.8266 & 0.5434 & 0.5262 & 0.6952 & 0.6327 & 0.3849 & 0.4230 & 0.5449 & 0.5673 \\
 & 5 & 0.6423 & 0.7131 & 0.6755 & 0.8393 & 0.5472 & 0.5354 & 0.6986 & 0.6491 & 0.3849 & 0.4257 & 0.5449 & 0.5741 \\
\addlinespace[4pt]
\multirow{5}{*}{2} & 1 & 0.6615 & 0.7101 & 0.6967 & 0.8386 & 0.5536 & 0.5141 & 0.7003 & 0.6264 & 0.4270 & 0.4323 & 0.5947 & 0.5708 \\
 & 2 & 0.6813 & 0.7463 & 0.7017 & 0.8654 & 0.5709 & 0.5492 & 0.7087 & 0.6738 & 0.4491 & 0.4457 & 0.6200 & 0.5779 \\
 & 3 & 0.6930 & 0.7650 & 0.7031 & 0.8807 & 0.5717 & 0.5758 & 0.7227 & 0.7109 & 0.4717 & 0.4644 & 0.6333 & 0.5977 \\
 & 4 & 0.7031 & 0.7681 & 0.7092 & 0.8863 & 0.5853 & 0.5870 & 0.7282 & 0.7175 & 0.4844 & 0.4674 & 0.6500 & 0.5941 \\
 & 5 & 0.7101 & 0.7756 & 0.7168 & 0.8898 & 0.5820 & 0.5913 & 0.7423 & 0.7342 & 0.4909 & 0.4645 & 0.6559 & 0.5880 \\
\addlinespace[4pt]
\multirow{5}{*}{3} & 1 & 0.7045 & 0.7483 & 0.7367 & 0.8606 & 0.5843 & 0.5641 & 0.7121 & 0.6882 & 0.4930 & 0.4553 & 0.6442 & 0.5817 \\
 & 2 & 0.7244 & 0.7779 & 0.7477 & 0.8828 & 0.5954 & 0.5891 & 0.7312 & 0.7260 & 0.5303 & 0.4800 & 0.6626 & 0.6007 \\
 & 3 & \cellcolor{gray!18}0.7466 & \cellcolor{gray!18}0.7944 & 0.7662 & 0.8994 & \cellcolor{gray!18}0.6103 & \cellcolor{gray!18}0.6067 & \cellcolor{gray!18}0.7491 & \cellcolor{gray!18}0.7537 & 0.5440 & 0.4946 & \cellcolor{gray!18}0.6767 & \cellcolor{gray!18}0.6170 \\
 & 4 & 0.7242 & 0.8014 & 0.7695 & 0.9068 & 0.6096 & 0.6126 & 0.7434 & 0.7597 & \cellcolor{gray!18}0.5430 & \cellcolor{gray!18}0.5010 & 0.6728 & 0.6229 \\
 & 5 & 0.7061 & 0.8086 & \cellcolor{gray!18}0.7897 & \cellcolor{gray!18}0.9096 & 0.5938 & 0.6130 & 0.7451 & 0.7698 & 0.5359 & 0.4957 & 0.6619 & 0.6116 \\
\bottomrule
\end{tabular}
\end{table*}

\subsubsection{Sensitivity to Ontology Mapping}
\label{app:ontology_mapping}
Table~\ref{tab:af3_results}  examines the sensitivity of our AF3 detector and mitigation evaluation to AudioSet ontology mapping. Let $z$ be a generated event label and $g$ a comparison label. \textbf{Depth=3} projects both labels to depth three and matches them if their projected sets overlap. Roots have depth one; shorter paths retain their deepest available node, and all paths are retained for labels with multiple parents. \textbf{Ancestor} accepts $g=z$ or $g$ being an ancestor of $z$; \textbf{Descendant} accepts $g=z$ or $g$ being a descendant of $z$. \textbf{Ancestor-or-descendant} accepts either relation, while \textbf{Exact} requires $g=z$. Except for depth projection, sharing an ancestor alone does not establish a match. We only vary the matching rule.

\begin{table*}[h]
\centering
\caption{Sensitivity to AudioSet ontology mapping for AF3 (Ours). Top: event-level detection with fixed ground-truth labels and detection parameters. Bottom: mitigation evaluation of fixed responses under different mapping rules. Detection metrics are on a 0--1 scale; mitigation metrics are percentages.}
\label{tab:af3_results}
\label{tab:af3_detection_mapping}
\small
\begin{tabular*}{\textwidth}{@{\extracolsep{\fill}}lrrrr@{}}
\toprule
\multicolumn{5}{l}{\textbf{Detection (Ours)}} \\
\midrule
\textbf{Mapping} & \textbf{Precision $\uparrow$} & \textbf{Recall $\uparrow$} & \textbf{F1 $\uparrow$} & \textbf{PR-AUC $\uparrow$} \\
\midrule
depth=3 & 0.7159 & 0.6517 & 0.6823 & 0.6799 \\
ancestor & 0.7888 & 0.6702 & 0.7247 & 0.7680 \\
descendant & 0.5924 & 0.7441 & 0.6596 & 0.6219 \\
ancestor-or-descendant & 0.8646 & 0.6570 & 0.7466 & 0.7944 \\
exact & 0.5538 & 0.7599 & 0.6407 & 0.5949 \\
\bottomrule
\end{tabular*}

\medskip
\begin{tabular*}{\textwidth}{@{\extracolsep{\fill}}lrrr@{}}
\toprule
\multicolumn{4}{l}{\textbf{Mitigation (Ours)}} \\
\midrule
\textbf{Mapping} & \textbf{ECHO-I $\downarrow$} & \textbf{ECHO-S $\downarrow$} & \textbf{Coverage $\uparrow$} \\
\midrule
depth=3 & 31.97 & 45.20 & 29.93 \\
ancestor & 35.95 & 50.40 & 28.21 \\
descendant & 19.81 & 30.80 & 33.41 \\
ancestor-or-descendant & 14.88 & 23.00 & 34.13 \\
\bottomrule
\end{tabular*}
\end{table*}

\subsection{Mitigation Ablations and Response Quality}
\label{app:mitigation_ablation}
We examine the masking budget and selection strategy, then evaluate the quality of the regenerated responses.

\subsubsection{Ablation on the Masking Budget}
\label{app:masking_budget}

We examine how the masking budget affects hallucination
mitigation. We vary $b \in \{1,5,10,15\}$ and set both
the initial selection size and the minimum masking
budget to $b$ (top-$b$/min-$b$).
For each predicted hallucinated event, we exclude
protected audio positions and replenish the selection
from the remaining attention ranking until $b$
positions are selected.
The detector configuration and all other settings
are kept fixed. The result is shown in Table~\ref{tab:masking_budget}.

\begin{table*}[h]
\centering\small
\caption{Effect of the differential masking budget $b$ on the 500-clip evaluation sets used in the main mitigation experiment. All metrics are percentages. Best and second-best results for each metric within each dataset--model group are shown in bold and underlined, respectively.}
\label{tab:masking_budget}
\setlength{\tabcolsep}{5pt}
\begin{tabular}{@{}lcccccccccccc@{}}
\toprule
& \multicolumn{4}{c}{ECHO-I $\downarrow$} & \multicolumn{4}{c}{ECHO-S $\downarrow$} & \multicolumn{4}{c}{Coverage $\uparrow$} \\
\cmidrule(lr){2-5}\cmidrule(lr){6-9}\cmidrule(lr){10-13}
Model & 1 & 5 & 10 & 15 & 1 & 5 & 10 & 15 & 1 & 5 & 10 & 15 \\
\midrule
\multicolumn{13}{@{}l}{\textbf{AudioSet Strong}} \\
Audio Flamingo 3 & 21.1 & \underline{18.1} & \textbf{14.9} & 18.5 & 32.4 & \underline{27.6} & \textbf{23.0} & 28.8 & 32.6 & \underline{33.9} & \textbf{34.1} & 33.1 \\
Qwen2.5-Omni & 26.7 & \underline{24.1} & \textbf{21.4} & 25.4 & 44.2 & \underline{42.2} & \textbf{37.0} & 42.6 & 36.9 & \underline{38.0} & \textbf{38.6} & 37.0 \\
Step-Audio 2 mini & 19.6 & \textbf{19.2} & \underline{19.4} & 19.6 & 30.6 & 29.4 & \textbf{28.6} & \underline{29.2} & \textbf{30.4} & \underline{29.6} & 27.8 & 28.6 \\
\midrule
\multicolumn{13}{@{}l}{\textbf{DCASE2019}} \\
Audio Flamingo 3 & 3.4 & 3.0 & \textbf{2.2} & \underline{2.7} & 4.0 & 3.6 & \textbf{2.6} & \underline{3.2} & \underline{60.7} & \textbf{60.9} & 60.3 & 59.8 \\
Qwen2.5-Omni & 7.1 & 7.7 & \textbf{6.5} & \underline{6.9} & 7.8 & 8.6 & \textbf{6.6} & \underline{7.4} & \textbf{58.6} & \underline{57.4} & 54.7 & 54.0 \\
Step-Audio 2 mini & 7.2 & \underline{6.6} & \textbf{6.4} & 6.9 & 8.6 & \underline{8.0} & \textbf{7.4} & 8.2 & \underline{61.2} & \textbf{61.3} & 58.3 & 58.7 \\
\bottomrule
\end{tabular}
\end{table*}

\subsubsection{Effect of Masking Strategy}
\label{app:masking_strategy}

We compare four masking strategies on AudioSet Strong using Audio Flamingo 3, Qwen2.5-Omni, and Step-Audio 2 mini, reusing the same 500 samples from the hallucination mitigation experiment for each model: \textbf{Contrast Masking} masks the audio positions most attended by events predicted as grounded, \textbf{Random Masking} selects audio positions uniformly at random, \textbf{Direct Masking} selects the ten audio positions most attended by each predicted hallucinated event, and \textbf{Differential Masking} excludes positions protected by predicted grounded events and replenishes the selection with the next highest-ranked unprotected positions. The results are shown in Table~\ref{tab:masking_strategy}.

\begin{table}[h]
    \centering
    \caption{Comparison of masking strategies on AudioSet Strong.
    All metrics are percentages. Bold indicates the best result
    within each model.}
    \label{tab:masking_strategy}
    \small
    \setlength{\tabcolsep}{6pt}
    \renewcommand{\arraystretch}{1.12}
    \begin{tabular}{@{}llrrr@{}}
        \toprule
        Model & Method
        & ECHO-I $\downarrow$
        & ECHO-S $\downarrow$
        & Coverage $\uparrow$ \\
        \midrule
        \multirow{4}{*}{Audio Flamingo 3}
        & Contrast Masking & 30.1 & 44.2 & 29.7 \\
        & Random Masking   & 26.7 & 39.8 & 31.6 \\
        & Direct Masking   & 16.8 & 25.0 & 33.5 \\
        & Differential Masking (Ours)
        & \textbf{14.9} & \textbf{23.0} & \textbf{34.1} \\
        \midrule
        \multirow{4}{*}{Qwen2.5-Omni-7B}
        & Contrast Masking & 43.0 & 66.4 & 31.4 \\
        & Random Masking   & 38.7 & 57.8 & 32.1 \\
        & Direct Masking   & 23.4 & 39.0 & 36.7 \\
        & Differential Masking (Ours)
        & \textbf{21.4} & \textbf{37.0} & \textbf{38.6} \\
        \midrule
        \multirow{4}{*}{Step-Audio 2 mini}
        & Contrast Masking & 40.5 & 59.8 & 21.5 \\
        & Random Masking   & 35.2 & 52.4 & 23.6 \\
        & Direct Masking   & \textbf{19.4} & 28.8 & \textbf{27.8} \\
        & Differential Masking (Ours)
        & \textbf{19.4} & \textbf{28.6} & \textbf{27.8} \\
        \bottomrule
    \end{tabular}
\end{table}

\subsubsection{Response Quality}
Following AudioCaps \citep{kim2019audiocaps}, we evaluate
response(caption) quality using BLEU-4, METEOR, ROUGE-L, and CIDEr on AudioCaps test set to validate that our method preserves the quality of the original responses. The results are shown in Table ~\ref{tab:caption_quality}.

\begin{table*}[h]
\centering
\caption{Response quality before and after masking on model-specific
subsets of the AudioCaps test set (we get 965/975 clips in the test set), restricted to samples
whose original responses contain hallucinated sound events. CLAP compares both responses with the original
audio; text metrics use all five reference descriptions.
All metrics are higher-is-better.}
\label{tab:caption_quality}
\setlength{\tabcolsep}{4pt}
\small
\begin{tabular}{lclccccc}
\toprule
Model & Original Hal.\ (\%) & Caption
& CLAP & BLEU-4 & METEOR & ROUGE-L & CIDEr \\
\midrule
\multirow{2}{*}{Audio Flamingo 3}
& \multirow{2}{*}{24.35}
& Original
& 0.4705 & 0.2303 & 0.2526 & 0.4893 & 0.4963 \\
& & After masking
& 0.4690 & 0.2785 & 0.2855 & 0.5172 & 0.6098 \\
\midrule
\multirow{2}{*}{Qwen2.5-Omni-7B}
& \multirow{2}{*}{41.35}
& Original
& 0.5259 & 0.0157 & 0.1409 & 0.1635 & 0.0004 \\
& & After masking
& 0.5248 & 0.0168 & 0.1447 & 0.1736 & 0.0022 \\
\midrule
\multirow{2}{*}{Step-Audio 2 mini}
& \multirow{2}{*}{19.79}
& Original
& 0.4473 & 0.0474 & 0.1760 & 0.3478 & 0.2948 \\
& & After masking
& 0.4626 & 0.0516 & 0.1828 & 0.3439 & 0.3333 \\
\bottomrule
\end{tabular}
\end{table*}

\subsection{Case Studies}
\label{app:qualitative_cases}

We present additional illustrative LatentLens decoding results and mitigation results produced by our method in the following: from top to bottom, we show the original response, the mel spectrogram of the original audio with ground-truth events, the attention heatmaps within 2 s around the most attended audio token positions for the two events, the LatentLens decoding results of the audio hidden states, and the mitigation results obtained by our method.

\begin{figure}[htbp]
\centering
\includegraphics[width=\linewidth,height=0.90\textheight,keepaspectratio]{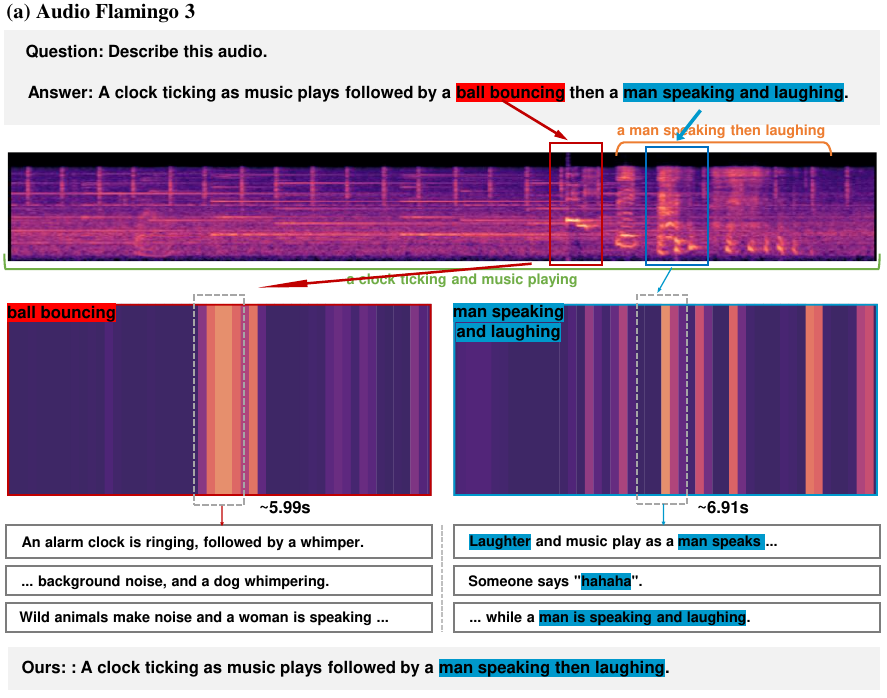}
\caption{Attention heatmaps, LatentLens decoding results, and hallucination mitigation for Audio Flamingo 3.}
\label{fig:app-heatmap}
\end{figure}

\begin{figure}[htbp]
\centering
\includegraphics[width=\linewidth,height=0.90\textheight,keepaspectratio]{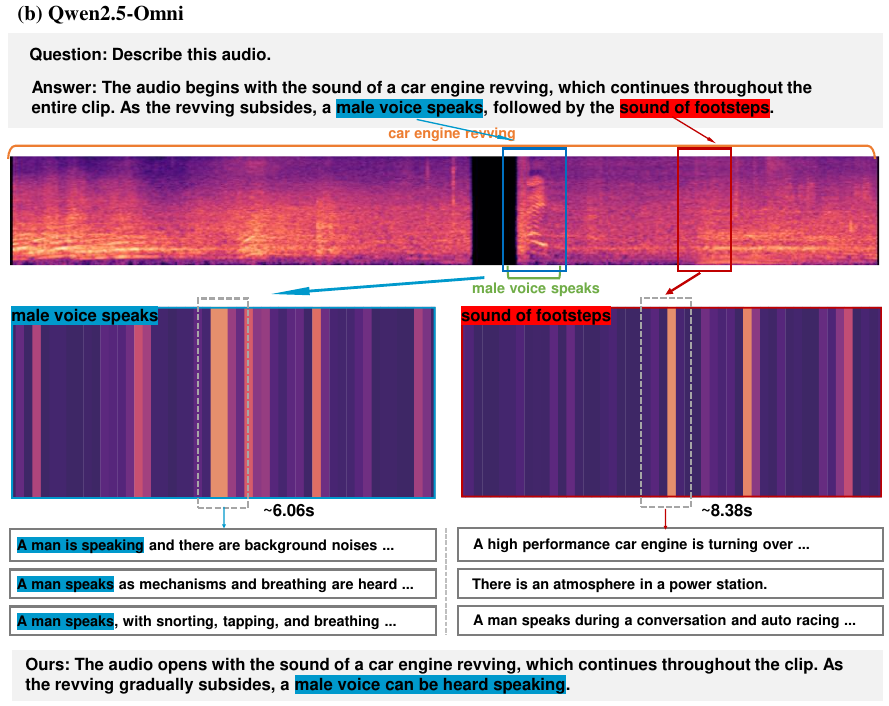}
\caption{Attention heatmaps, LatentLens decoding results, and hallucination mitigation for Qwen2.5-Omni.}
\label{fig:app-heatmap-omni}
\end{figure}
\clearpage

\begin{figure}[htbp]
\centering
\includegraphics[width=\linewidth,height=0.90\textheight,keepaspectratio]{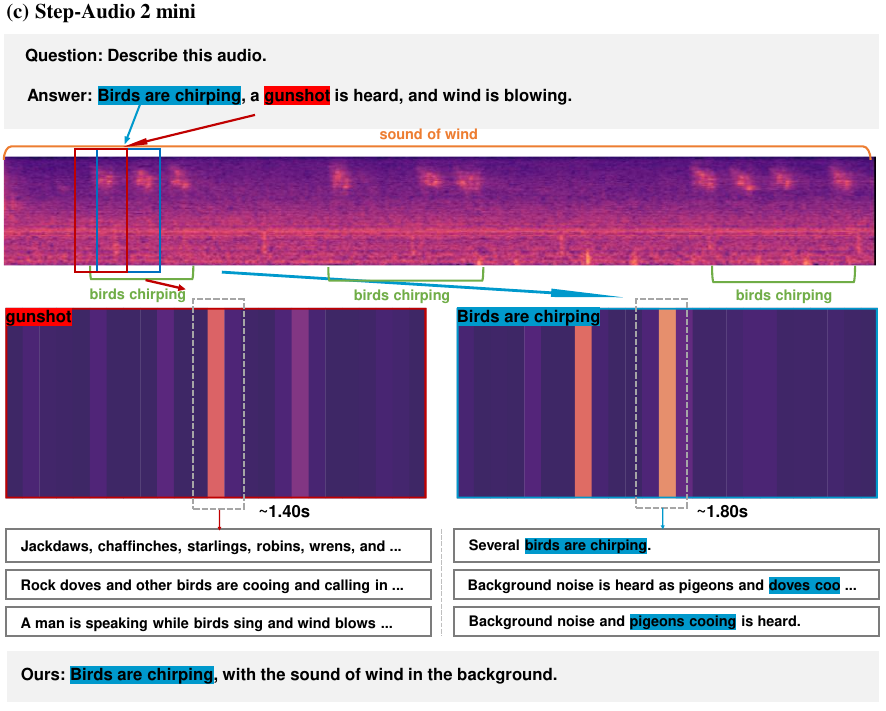}
\caption{Attention heatmaps, LatentLens decoding results, and hallucination mitigation for Step-Audio 2 mini.}
\label{fig:app-heatmap-step}
\end{figure}
\clearpage
\end{document}